\documentclass[acmsmall]{acmart}
\AtBeginDocument{%
  \providecommand\BibTeX{{%
    \normalfont B\kern-0.5em{\scshape i\kern-0.25em b}\kern-0.8em\TeX}}}

\setcopyright{acmcopyright}
\copyrightyear{2023}
\acmYear{2023}
\acmDOI{10.1145/1122445.1122456}

\acmJournal{TOSEM}
\acmVolume{37}
\acmNumber{4}
\acmArticle{111}
\acmMonth{11}

\usepackage{microtype}
\usepackage{textcomp}
\usepackage{bm}
\usepackage[caption=false]{subfig}
\usepackage{flushend}
\usepackage{csquotes}
\usepackage{booktabs} 
\usepackage{framed}
\usepackage{graphicx}
\usepackage{tabularx}
\usepackage{longtable}
\usepackage{graphicx}
\usepackage{url}
\usepackage{multirow}
\usepackage{enumitem}
\usepackage{hyperref}
\usepackage{lscape}
\usepackage[normalem]{ulem}
\useunder{\uline}{\ul}{}
\usepackage{latexsym}
\usepackage{xcolor}
\usepackage{color, colortbl}
\usepackage{hyperref}

\definecolor{g}{gray}{0.925}
\usepackage{graphics}

\definecolor{lightgray}{gray}{0.9}
 
\usepackage{siunitx}  
\usepackage{textgreek}  
\usepackage{wasysym}
\usepackage{makecell}

\usepackage{tabularx}
\begin{document}

\title[Understanding Developers Well-Being and Productivity]{Understanding Developers Well-Being and Productivity: a 2-year Longitudinal Analysis during the COVID-19 Pandemic}

\author{Daniel Russo}
\authornote{Corresponding author.}
\email{daniel.russo@cs.aau.dk}
\orcid{0000-0001-7253-101X}
\affiliation{%
  \institution{Department of Computer Science, Aalborg University}
  \streetaddress{A.C. Meyers Vaenge, 15, 2450}
  \city{Copenhagen}
  \country{Denmark}}

\author{Paul H. P. Hanel}
\email{p.hanel@essex.ac.uk}
\orcid{0000-0002-3225-1395}
\affiliation{%
  \institution{Department of Psychology, University of Essex}
  \country{United Kingdom}
}

\author{Niels van Berkel}
\email{nielsvanberkel@cs.aau.dk}
\orcid{0000-0001-5106-7692}
\affiliation{%
  \institution{Department of Computer Science, Aalborg University}
  \city{Aalborg}
  \country{Denmark}
  }

\renewcommand{\shortauthors}{Russo et al., 2023}

\begin{abstract}

The COVID-19 pandemic has brought significant and enduring shifts in various aspects of life, including increased flexibility in work arrangements. In a longitudinal study, spanning 24 months with six measurement points from April 2020 to April 2022, we explore changes in well-being, productivity, social contacts, and needs of software engineers during this time. Our findings indicate systematic changes in various variables. For example, well-being and quality of social contacts increased while emotional loneliness decreased as lockdown measures were relaxed. Conversely, people's boredom and productivity, remained stable. Furthermore, a preliminary investigation into the future of work at the end of the pandemic revealed a consensus among developers for a preference of hybrid work arrangements. We also discovered that prior job changes and low job satisfaction were consistently linked to intentions to change jobs if current work conditions do not meet developers' needs. This highlights the need for software organizations to adapt to various work arrangements to remain competitive employers. Building upon our findings and the existing literature, we introduce the Integrated Job Demands-Resources and Self-Determination (IJARS) Model as a comprehensive framework to explain the well-being and productivity of software engineers during the COVID-19 pandemic. 

\end{abstract}



\maketitle

\section{Introduction}

The COVID-19 pandemic and the subsequent lockdowns are likely among the most disruptive events that most software engineers in Western countries faced during their lifetime.
Suddenly, professionals started to work from home, potentially alongside family members.
This peculiar situation is unprecedented in computer science history; thus, we need more information about the long-lasting impact of lockdowns on the well-being and productivity of software professionals.

The only related evidence comes from the effects of quarantined people in previous epidemic outbreaks, which suggests that isolation and lockdown measures are a burden to individuals' well-being~\cite{brooks2020,de2023subjective,gutierrez2022face} and productivity~\cite{lipsitch2020defining}. 
Indeed, well-being and productivity are two crucial aspects of our lives, particularly during extraordinary events---with well-being considered a fundamental right according to the Universal Declaration of Human Rights.
Health professionals have already identified some relevant predictors of well-being during harmful events~\cite{brooks2020,farmer1986boredom}.
However, this research is often cross-sectional (i.e., not longitudinal), only includes a limited number of well-being-related variables, and focuses on well-being while ignoring productivity.
The software engineering community reacted quickly to this event by conducting an extensive study that found that home office ergonomics, disaster preparedness, and fear correlate with well-being and productivity~\cite{Ralph2020pandemic}.
However, Ralph et al. performed a cross-sectional study with only a few predictors.
Pre-pandemic research on remote work~\cite{donnelly2015disrupted} might provide some indications.
However, such research is unlikely to be relevant during a global pandemic, with professionals locked down in their houses without childcare or usual welfare support provided during non-pandemic times.

For these reasons, it is essential to continuously and longitudinally investigate software professionals' well-being and productivity across the COVID-19 pandemic (until spring 2022).
By doing so, we aimed to achieve several goals.
First, test whether well-being, productivity, and other relevant social and psychological variables related to emotions, personality, and work changed over the course of 24 months since the beginning of the first lockdown in Spring 2020.
Second, test whether well-being and productivity changed more for software developers who had experience in working from home, were living alone; and explore whether income, gender, age, and preference for working from home further interact with well-being and productivity over time.
Third, turning to the future, investigating predictors of job satisfaction and intention to change jobs.
Fourth, understand how to improve developers' work-life balance while working from home in a post-pandemic setting and contribute to the nascent literature about the future of work.
Hence, we formulate our research questions as follows:
\\[.1in]
\noindent\textit{\textbf{Research Question 1}}: \textit{
How and why have well-being, productivity, and other relevant social and psychological variables related to emotions, personality, and work changed during the COVID-19 pandemic?}
\\[.1in]
\noindent\textit{\textbf{Research Question 2}}: \textit{
How do software developers envision their future of work following the COVID-19 pandemic?
}
\\

To answer our research questions, we sampled almost 200 globally distributed software engineers six times over a period of 24 months. We assessed their well-being and productivity in each of the six waves alongside 15 other variables.
To guide our research design, we grounded our investigation in organizational~\cite{herzberg2017motivation} and psychological~\cite{ryan2000self} theories, which are relevant to people's well-being and productivity. For example, self-determination theory~\cite{ryan2000self} assumes that human motivation can be divided into three basic needs, which are also linked with work motivation~\cite{gagne2005self}: the needs for autonomy, competence, and relatedness.   
Additionally, we also included evidence from the remote work literature~\cite{Lascau2019workers,anderson2015impact,bloom2015does}, and recommendations by health and work authorities~\cite{nhs_2020,sst_2020,CIPD}.

This investigation is framed within an overarching research agenda regarding the Future of Work in the software industry.
In particular, this work aims to monitor the effects on developers' well-being and productivity during two years of the COVID-19 pandemic with a longitudinal design. 
In these years, we also published preliminary findings to provide early recommendations to professionals and organizations~\cite{russo2020predictors}.
Additionally, we also looked more specifically into daily life practices of locked down software engineers~\cite{russo2021daily} while also monitoring how different activities (e.g., bugfixing, coding, helping) impacts software engineers' satisfactions and performance~\cite{russo2021developers}. 
Finally, we also investigated how the pandemic influenced Agile development methodologies, particularly Scrum~\cite{Cucolas2021Scrum,verwijs2021theory}. 

In this study, we analyzed our data using a range of different statistical approaches tailored to the specific questions. Specifically, to test whether well-being, productivity, and 15 variables, including loneliness, needs, and social contacts changed over time, we used 17 Friedman's tests. We used a series of linear mixed-effects models to test whether the changes differed across people (e.g., for those with more WFH experience). To assess whether there are any mean differences between participants living in the UK and USA, we used a series of between-subject t-tests. 
We found that software engineers' levels of well-being increased while emotional loneliness decreased between April 2020 and April 2022. Productivity remained unchanged. Additionally, we found that having previously changed jobs and low job satisfaction was reliably associated with the intention to change jobs again. Further, especially the need for competence and solidarity with the company was positively associated with job satisfaction. Moreover, the thematic analysis revealed that software engineers perceive hybrid work as the new norm in the tech industry. Finally, we found no mean differences between people living in the UK and USA for any of the 17 variables we measured across all six waves. 
Building on the findings of this study and previous research, we present the Integrated Job Demands-Resources and Self-Determination (IJARS) Model as an all-encompassing structure for understanding the well-being and efficiency of software engineers amidst the COVID-19 crisis. The IJARS model fuses components from the Job Demands-Resources (JD-R) Model and Self-Determination Theory (SDT) to investigate the interplay between job demands and resources in relation to fulfilling fundamental psychological necessities (autonomy, competence, and relatedness) for remote work.

This article has the following structure.
Section~\ref{sec:related} discusses related work of well-being and productivity.
The research design and analysis are then described in Section~\ref{sec:design}.
Next, in Section~\ref{sec:results}, we discuss the results of our analyses, while we discuss implications and recommendations for professionals and software companies in Section~\ref{sec:discussion}.
The Integrated Job Demands-Resources and Self-Determination (IJARS) Model is then explained in Section~\ref{sec:theory}.
Finally, we conclude our work by outlying future research directions in Section~\ref{sec:conclusion}.

\section{Related Work}
\label{sec:related}
Following the abrupt onset of the COVID-19 pandemic and subsequent lockdowns, COVID-19-related research has expanded rapidly.
Health scientists started investigating countermeasures to reduce the spread and impact of the virus and studied the psychological and physiological effects on people living in lockdown conditions.
Also, in the software engineering community, the effect of the pandemic on software developers has gained increased attention.
After describing the state of the art of research on well-being and productivity in remote work, we focus on the software engineering contributions.

\subsection{Well-Being and Productivity in Remote Work}

There is a consensus that lockdown measures have a negative impact on well-being~\cite{brooks2020,foad2021limitations,lunn2020using}. 
In particular, research shows that living in a lockdown can result in increased experiences of anger, depression, emotional exhaustion, fear of infecting others or getting infected, insomnia, irritability, loneliness, low mood, post-traumatic stress disorders, and stress~\cite{sprang2013posttraumatic,hawryluck2004sars,lee2005experience,marjanovic2007relevance,reynolds2008understanding,bai2004survey}. 
Additionally, fears of, e.g., infection~\cite{kim2015public,prati2011social}, lack of supplies or not being treated~\cite{wilken2017knowledge}, and misleading or contradictory information~\cite{caleo2018factors} can result in significantly increased stress levels.
Moreover, the psychological effects of being locked down may appear years after~\cite{brooks2020}. 

On the other hand, pre-COVID research shows that remote working is associated with an improved work-life balance, creativity, productivity, reduced stress, and low carbon emissions due to the absence of commuting~\cite{owl_labs_2019,anderson2015impact,bloom2015does,vega2015within,baruch2000teleworking,cascio2000managing}, even though recent studies challenge the claim that remote work leads to lower carbon emissions~\cite{caldarola2022teleworkers,wohner2022work}.
Nevertheless, there are also some apparent drawbacks related to remote work, such as deteriorating collaboration and communication, loneliness, feeling of being constantly `online,' decreasing motivation, and distractions at home~\cite{buffer2020,knight2022loneliness}. 
Independent of whether the positive effects outweigh the negative effects, forecasts suggest that remote work will increase on a large scale in the next years~\cite{owl_labs_2019,gallup2020}.

\subsection{Software Engineering and Remote Work}
Overall, the software engineering community has been quite active in researching remote work, already before the onset of the COVID-19 pandemic.
We identified relevant work through Scopus.

The first works in this research area are from the late 90s with broader internet use.
Pounder~\cite{pounder1998homeworking} was the first relevant contribution we identified, with an essay about security problems linked to telework.
In the early 2000s, Guo~\cite{guo2001special} performed two qualitative surveys on software process improvement related to the distinctive nature of teleworking. Similarly, Higa et al.~\cite{higa2000understanding} studied how e-mail usage influences telework. 

Afterward, there has been a lacking interest on the topic by the software engineering community, with only two exceptions.
James \& Griffiths~\cite{james2014secure} developed a mobile execution
environment to support a secure and portable working from home setting. Ford et al.~\cite{ford2019remote} interviewed three transgender software engineers to explore the interplay of gender identity and remote work. 

Following the start of the COVID-19 pandemic and the first lockdown, two research groups performed survey studies.
Ralph et al.~\cite{Ralph2020pandemic} performed a cross-sectional study of over two thousand globally distributed developers working from home during the pandemic where an \textit{a priori} research model derived by literature was validated through Structural Equation Modeling.
Russo et al.~\cite{russo2020predictors} went in the opposite direction. Rather than having a top-down model to validate, they employed an exploratory approach investigating the most relevant variables related to either well-being or productivity. They used a longitudinal design.
 
Microsoft has also been active in understanding the effects of the pandemic on its employees.
Ford et al.~\cite{ford2020tale} surveyed Microsoft's developers twice.
They found that the quality of family life and time improved, although remote work introduced a lack of focus, poor work-life boundaries, and communication and sync issues.
Similarly, Miller et al.~\cite{miller2021your} performed two surveys in which they collected information about working from home and team-related issues. They found that communication and colleague interaction are relevant predictors of developers' satisfaction and team productivity. 
Butler \& Jaffe~\cite{butler2021challenges} conducted a 10-week diary study. Identified challenges from remote work were meetings, overwork, and physical and mental health. However, Microsoft developers appreciated more family time and work flexibility. 

More recent studies focus on particular aspects of remote work.
For example, Cucolaș \& Russo~\cite{Cucolas2021Scrum}, with a Mixed-Methods research design, investigated how Scrum software development adapted to working from home.
According to their results, the home-working environment is the most crucial variable for a software project's success. Also, self-determination theory~\cite{ryan2000self} (i.e., the need for autonomy, competence, and relatedness) is a valuable theoretical lens to improve working from home conditions, as they are linked with well-being~\cite{cantarero2021affirming}, for example.
Machado et al.~\cite{machado2021gendered} surveyed 233 Brazilian software professionals and investigated gender differences.
They concluded that the pandemic affected women more negatively than men.
In contrast, Russo et al. did not find any meaningful gender differences~\cite{russo2020predictors}.
Documentation and setup in the initial months of the lockdown have also been perceived as especially struggling by developers, according  to~\cite{uddin2022qualitative}. 
These authors claim for broader use of automated tools for remote work.
Finally, consistent with the findings from Russo et al., \v{S}mite
 et al. confirm that, on average, perceived productivity has not significantly changed during the pandemic~\cite{smite2022changes}.

From a content perspective, approximately half of the papers are concerned with specific topics related to remote work: job security~\cite{pounder1998homeworking,james2014secure,uddin2022qualitative}, process~\cite{guo2001special}, work productivity~\cite{higa2000understanding,smite2022changes}, and inclusion~\cite{ford2019remote}.
The other half of the papers focused on the well-being and productivity aspects of remote work~\cite{ford2020tale,Ralph2020pandemic,russo2020predictors,butler2021challenges,machado2021gendered,lamarche2020socially} and productivity related to project characteristics~\cite{Cucolas2021Scrum}.

\section{Research Design}
\label{sec:design}

To design our research, we followed the ACM SIGSOFT Empirical Standards for Longitudinal Studies~\cite{ralph2020empirical}.
Consequently, we asked carefully recruited software professionals to complete the same survey six times, over a period of 24 months. 

Due to the unpredictable evolution of the pandemic and state-prescribed lockdown regulations we could not plan data collection beforehand. 
Thus, we tried to follow the evolution of the pandemic.
Wave 1 was collected between 26-30 April 2020, wave 2 between 10-13 May 2020, wave 3 between 24 February and 3 March 2021, wave 4 between 29 June and 5 July 2021, wave 5 between 20 December 2021 and 5 January 2022, and wave 6 between 6-12 April 2022. Wave 1 and 2 were only two weeks apart since we were initially only interested in the stability of predictors of well-being and productivity.
Wave 3 was collected in late winter 2021 when the number of COVID-19 cases in most Western countries decreased again. Wave 4 was collected when a significant part of people in Western countries had received an offer to get vaccinated, wave 5 around Christmas 2021 when a new Covid-wave hit many countries, and wave 6 when cases dropped again in spring 2022.  
Unique randomized IDs were assigned to participants to preserve their anonymity and track their participation across all six waves. 

Our research employed a longitudinal design, which was selected for its ability to capture changes over time. As Rindfleisch et al. have noted, even cross-sectional studies that are carefully designed cannot reveal how phenomena evolve over time~\cite{rindfleisch2008cross}. By tracking the same group of participants over an extended period, we were able to examine the changes that occurred in our variables of interest and explore underlying mechanisms. 
Therefore, we believe that our choice of a longitudinal design was essential for achieving the aims of our study.

Additionally to investigating changes over time in well-being, productivity, as well as well-being related variables (e.g., anxiety, loneliness, stress), needs, and coping strategies, we were interested in predictors of well-being and productivity. In wave 1, we included over 50 variables that were previously linked to well-being or productivity (for more details see~\cite{russo2020predictors}). Only the variables which were reliably associated with well-being or productivity were included in the subsequent waves. For example, we predicted based on self-determination theory~\cite{ryan2000self} that the needs for autonomy, competence, and relatedness are positively associated with well-being and productivity~\cite{russo2020predictors}. Further, in line with other research we expected that anxiety, loneliness, and stress are negatively associated with well-being~\cite{Spitzer2006GAD7, Cohen1988perceived, gierveld2006}, whereas resilience and quality of social contacts were expected to be positively associated with well-being~\cite{smith2008brief}.  

Similarly, we also expected that the needs for autonomy, competence, and relatedness are positively associated with productivity~\cite{baard2004intrinsic}. 


\subsection{Participants}
We selected participants from a pool of 483 software engineers previously identified~\cite{russo2020gender}.
To ensure high-quality data, in Russo \& Stol we implemented a three-step screening process for participants in our data collection platform. In the first step, we \textit{pre-screened} potential members based on several criteria, including their knowledge of software development techniques, their profession as computer programmers, their use of technology at work, and their reliability score of 100\% on the Prolific platform. Out of 75,296 Prolific members active in the last three months, we included 2,897 members who met our pre-screening criteria.

In the second step, we conducted a \textit{competence screening} by sending a screening survey to a random subset of the 2,897 potential subjects until we reached around 1,000 willing participants. Only those who self-identified as software professionals and demonstrated an adequate level of knowledge of software development through a competency-based questionnaire were invited to participate in our study. 
In particular, we asked three multiple-choice questions,\footnote{Questions are included in the Appendix~\ref{sec:appendixCompetence}.} one about software design and two about programming, and also time-boxed the responses within three minutes to avoid suspicious behavior.
This step resulted in 514 eligible candidates, with 154 being excluded for incorrect answers and 92 for taking too long to complete the survey.

In the final step, we implemented \textit{quality screening} by including attention checks in the full questionnaire and excluding eight participants who failed to recognize them. The questions were randomized within blocks to minimize response bias. After completing the screening process, we included 483 valid and complete responses.

For this investigation, we narrowed this initial pool down to 192 professionals through additional screening questions.
In particular, informants should not have lived in countries with non-uniform COVID regulations (e.g., Germany, where initially the lockdown regulations were provided by the regions, not the federal government). In addition, we looked for informants working from home during the pandemic for at least 50\% of their time. 192 software engineers completed the first survey ($M_{age}$~=~36.65 years, $SD$~=~10.77, range~=~19–63; 154 men, 38 women), 184 participated in wave 2, 144 in wave 3, 125 in wave 4, 117 in wave 5, and 101 in wave 6. 
Overall, 72 participated in all six waves and completed all measures. A sensitivity analysis with GPower 3.1.9.4~\cite{faul2009statistical} revealed that this sample size was sufficient to detect a medium effect size of $f = .15$ with a power of .95.

Demographic information is provided in Table~\ref{tab:demographicsoverview}. Twenty-nine participants were living alone at wave 1, 162 with other people. 
We recruited participants from the academic data collection platform Prolific~\cite{palan2018prolific} and compensated participants above the US's federal minimum wage\footnote{Payments slightly changed during the years to adapt to new labor regulations and Prolific recommendations. We always used the suggested compensation by Prolific. As a reference, we paid our participants GBP 9.00/hour in the last wave.}.
This compensation was used for all participants.
Additionally, none of our participants failed any attention checks or completed the survey too fast, ensuring our data quality.
The survey was run using the platform Qualtrics.

\begin{table}[]
\caption{Overview of participant demographics across the six waves.}
\centering
\begin{tabular}{@{}llllllll@{}}
\toprule
 & \textit{\textbf{N}} & \textbf{Men} & \textbf{Women} & \textbf{Age (mean)} \\ \midrule
Wave 1 & 192 & 154 & 38 & 36.65 \\
Wave 2 & 184 & 147 & 37 & 36.71 \\
Wave 3 & 144 & 112 & 27 & 37.56 \\
Wave 4 & 125 & 96 & 27 & 39.20 \\
Wave 5 & 117 & 91 & 22 & 40.12 \\
Wave 6 & 101 & 81  & 16 & 41.16 \\ \bottomrule
\end{tabular}
\label{tab:demographicsoverview}
\end{table}

We followed the ethical guidelines of the Declaration of Helsinki throughout data collection and analysis~\cite{general2014Helsinki}.
All participants were at least 18 years old, expressed their consent to participate in the study each time, and were free to withdraw at any point.
All the authors completed formal training in research ethics for engineering and behavioral sciences.

\subsection{Measurements}
\label{sec:measurements}

Well-being and productivity are two complementary variables of a healthy working environment.
Not surprisingly, they are correlated, and greater happiness can cause greater productivity~\cite{oswald2015happiness, russo2020predictors}.
Especially in exceptional times, such as a pandemic, organizations should prioritize employees' mental and physical well-being if they want to be productive. 
On the other hand, as suggested by Russo et al.~\cite{russo2020predictors}, contributing to the organization's value is essential for the sense of belonging or achievement of every developer.
Therefore, productivity also contributes to professionals' well-being \cite{russo2020predictors}.

Consequently, productivity and well-being are our two outcome variables (i.e., dependent variables).
To identify relevant predictors (or our independent variables) of our dependent variables, we started from the insights of Russo et al.~\cite{russo2020predictors}. 
Namely, we included in this analysis only the 15 (out of 50) predictors which correlated with at least one of the two outcome variables (i.e., $r {\geq}  |.30|$)~\cite{russo2020predictors}. This was done to keep the number of predictor variables to a manageable amount and to focus on the most relevant variables. 

All variables were measured using self-reported measures, which is very common in the literature~\cite{Ralph2020pandemic,russo2020gender}. The internal consistency of the scales was quantified with Cronbach's $\alpha$ and ranged from satisfactory to very good. Values above .60 and .70 are desirable for exploratory and confirmatory research, respectively~\cite{hair2013multivariate}. We created the composite variables by averaging all items of a scale and after reverse scoring (recoding) some items of some scales (if applicable). This also ensured that each scale had (often substantially) more than five response categories, so they can be treated as a continuous variable~\cite{rhemtulla2012can}.

To measure the identified variables, we only used either validated scales or adapted items from scales used in previous publications with high reliabilities. The only exceptions were `productivity,' `quality and quantity of communication with colleagues and line managers, and `daily routines' for which we created our own items because we could not find existing scales suitable for our purposes. 
Responses were mainly given on 5-, 6-, or 7-point response scales, with higher values indicating a higher score on each variable.
Every scale is briefly subsequently described with its name, reference, and reliability metrics (i.e., Cronbach's alpha) across all six data collection waves. For detailed descriptions of the items, see Appendix~\ref{sec:appendixQuestions} and Appendix~\ref{sec:appendixQuestionsW6}.

\textbf{Well-being}. We measured well-being with the 5-item Satisfaction with Life Scale~\cite{diener1985satisfaction}. Participants were asked to report their well-being using items such as "I was satisfied with my life in the past week" on a 7-point Likert scale (1: Strongly disagree, 7: Strongly agree). The Cronbach's $\alpha$ values to measure internal consistency for all six data collection waves were the following $\alpha_{Wave 1}~=~.90$, $\alpha_{Wave 2}~=~.90$, $\alpha_{Wave 3}~=~.92$, $\alpha_{Wave 4}~=~.94$, $\alpha_{Wave 5}~=~.94$, $\alpha_{Wave 6}~=~.95$. 

\textbf{Productivity}. There is no agreement among researchers on how productivity can be measured. For example, measuring productivity in an allegedly objective way by using function points~\cite{wagner2018systematic} has been criticized as detrimental in the long run~\cite{ko2019we}. Further, the objective approach is barely feasible if participants work in different areas since comparisons across work are very challenging. 
Therefore, other researchers advocated using self-reports~\cite{meyer2014software}, which has apparent shortcomings such as subjectivity.
In the present research, we developed a subjective approach. We reduce socially desirable responses by making the survey anonymous. 
Specifically, we operationalized productivity as a function of time spent working and efficiency per hour, compared to a typical, pre-pandemic week.
This choice is because we wanted to investigate productivity while working remotely compared to being in the office. Since our measure does not allow us to compute internal consistency, we instead computed test-retest reliability by correlating the productivity scores at Wave 1 with those at time t2 ($r_{it} = .50, p < .001$). 

\textbf{Boredom} was measured with the Boredom Proneness Scale~\cite{farmer1986boredom,struk2017short}; $\alpha_1~=~.87$, $\alpha_2~=~.87$, $\alpha_{Wave 3}~=~.92$, $\alpha_{Wave 4}~=~.90$, $\alpha_{Wave 5}~=~.92$, $\alpha_{Wave 6}~=~.91$.

\textbf{Self-blame and behavioral disengagement}, two coping strategies, were measured with the respective subdimensions of the Brief COPE scale~\cite{Carver1997BriefCOPE}. Cronbach's $\alpha$'s for self-blame were $\alpha_1~=~.75$, $\alpha_2~=~.71$, $\alpha_{Wave 3}~=~.92$, $\alpha_{Wave 4}~=~.92$, $\alpha_{Wave 5}~=~.88$, $\alpha_{Wave 6}~=~.90$, and for behavioral disengagement $\alpha_1~=~.76$, $\alpha_2~=~.71$, $\alpha_{Wave 3}~=~.89$, $\alpha_{Wave 4}~=~.91$, $\alpha_{Wave 5}~=~.95$, $\alpha_{Wave 6}~=~.92$.

\textbf{Distractions at home} was measured with a 2-item scale we developed ($\alpha_1~=~.64$, $\alpha_2~=~.63$, $\alpha_{Wave 3}~=~.75$, $\alpha_{Wave 4}~=~.65$, $\alpha_{Wave 5}~=~.65$, $\alpha_{Wave 6}~=~.63$. 

\textbf{Generalized anxiety} was measured with an adapted version of the $7$-item Generalized Anxiety Disorder scale~\cite{Spitzer2006GAD7}; $\alpha_1~=~.93$, $\alpha_2~=~.93$, $\alpha_{Wave 3}~=~.94$, $\alpha_{Wave 4}~=~.95$, $\alpha_{Wave 5}~=~.93$, $\alpha_{Wave 6}~=~.93$.

\textbf{Emotional and social loneliness} were measured with the De Jong Gierveld Loneliness Scale~\cite{gierveld2006}. Emotional loneliness' Cronbach's $\alpha$s were: $\alpha_1~=~.68$, $\alpha_2~=~.69$,  $\alpha_{Wave 3}~=~.68$, $\alpha_{Wave 4}~=~.73$, $\alpha_{Wave 5}~=~.70$, $\alpha_{Wave 6}~=~.69$, and for social loneliness: $\alpha_1~=~.84$, $\alpha_2~=~.87$, $\alpha_{Wave 3}~=~.90$, $\alpha_{Wave 4}~=~.88$, $\alpha_{Wave 5}~=~.91$, $\alpha_{Wave 6}~=~.94$. 

\textbf{Autonomy, competence, and relatedness} were measured with the psychological needs scale~\cite{sheldon2012balanced}. Need for autonomy's Cronbach's $\alpha$s were: $\alpha_1~=~.72$, $\alpha_2~=~.76$, $\alpha_{Wave 3}~=~.77$, $\alpha_{Wave 4}~=~.78$, $\alpha_{Wave 5}~=~.77$, $\alpha_{Wave 6}~=~.79$; for Competence: $\alpha_1~=~.77$, $\alpha_2~=~.65$, $\alpha_{Wave 3}~=~.77$, $\alpha_{Wave 4}~=~.79$, $\alpha_{Wave 5}~=~.77$, $\alpha_{Wave 6}~=~.84$; and for Relatedness: $\alpha_1~=~.79$, $\alpha_2~=~.78$, $\alpha_{Wave 3}~=~.78$, $\alpha_{Wave 4}~=~.80$, $\alpha_{Wave 5}~=~.77$, $\alpha_{Wave 6}~=~.80$.

\textbf{Quality of social contacts} were measured with 3-items, two of which were adapted from the social relationship quality scale~\cite{birditt2007relationship} and one was developed by us, $\alpha_1~=~.73$, $\alpha_2~=~.77$, $\alpha_{Wave 3}~=~.76$, $\alpha_{Wave 4}~=~.84$, $\alpha_{Wave 5}~=~.80$, $\alpha_{Wave 6}~=~.85$.  

\textbf{Quality and quantity of communication with colleagues and line managers} were measured with a self-developed 3-item scale, $\alpha_1~=~.88$, $\alpha_2~=~.92$, $\alpha_{Wave 3}~=~.93$, $\alpha_{Wave 4}~=~.94$, $\alpha_{Wave 5}~=~.94$, $\alpha_{Wave 6}~=~.93$.

\textbf{Stress} was measured with the Perceived Stress Scale~\cite{cohen1983global}; $\alpha_1~=~.80$, $\alpha_2~=~.77$, $\alpha_{Wave 3}~=~.83$, $\alpha_{Wave 4}~=~.78$, $\alpha_{Wave 5}~=~.76$, $\alpha_{Wave 6}~=~.80$.

\textbf{Daily Routines} were measured by a self-developed 5-item scale ($\alpha_1~=~.75$, $\alpha_2~=~.78$, $\alpha_{Wave 3}~=~.81$, $\alpha_{Wave 4}~=~.78$, $\alpha_{Wave 5}~=~.82$, $\alpha_{Wave 6}~=~.79$.

\textbf{Extraversion} was measured with a subscale of the Brief HEXACO Inventory~\cite{DeVries2013HEXACO}; $\alpha_1~=~.71$, $\alpha_2~=~.69$, $\alpha_{Wave 3}~=~.75$, $\alpha_{Wave 4}~=~.61$, $\alpha_{Wave 5}~=~.68$, $\alpha_{Wave 6}~=~.73$.

Additionally, to answer Research Question 2, we included the following measures in wave 6.

\textbf{Preference for working from home} was measured with a 2-item slider scale we developed with responses ranging from 0 to 100, $\alpha_{Wave 6}~=~.88$.

\textbf{Resilience} was measured with the 6-item Brief Resilience Scale~\cite{smith2008brief}, $\alpha_{Wave 6}~=~.89$. 

\textbf{Solidarity with the company} was measured with the 3-item Solidarity subscale of the In-group Identification Scale~\cite{leach2008group}, $\alpha_{Wave 6}~=~.97$.

\textbf{Job satisfaction} was measured with the 4 highest loading items of the Generic Job Satisfaction Scale~\cite{macdonald1997generic}, $\alpha_{Wave 6}~=~.85$.

\textbf{Disliking commuting} was measured with a 2-item scale we developed, $\alpha_{Wave 6}~=~.54$.

\textbf{Intention to change jobs} was measured with a 2-item scale we developed, $\alpha_{Wave 6}~=~.95$

\textbf{Changed jobs} since March 2020 was measured with one item we developed whereby 0 represents no and 1 yes. 

\textbf{Perceived company preference for WFH} was measured with a one item slider scale ranging from 0 (company wants me to go back to the office full-time) to 100 (company allows me to work from anywhere full-time).

\subsection{Analysis}

To answer our research question and perform the additional exploratory analysis, we used various statistical analyses. Below, we briefly describe and justify each of them. Given our rich dataset, we corrected for multiple comparison to reduce the risk of a false-positive finding to the default level of $\alpha = .05$. A common method is to divide .05 by the number of tests -- the so-called Bonferroni correction. However, this method is too conservative and thus reduces statistical power, especially if variables are correlated. We therefore used the $M_{eff}$-correction method~\cite{derringer2018simple}, which originates from genetic research~\cite{cheverud2001simple, nyholt2004simple} and takes dependency into account. If all variables are uncorrelated, this approach is identical to a Bonferroni correction. Conversely, if the correlations between all variables approach 1, the corrected $\alpha$-threshold approaches .05.  

Raw data, R-code to reproduce our analyses, and the zero-order correlations for all 17 variables, separately per wave and across all six waves, are included in our replication package on Zenodo\footnote{Link to the replication package: https://zenodo.org/record/7828605.}.

\subsubsection{Changes along the COVID-19 Pandemic}
\label{sec:wANOVA}
To test whether any change between the six data collection waves occurred, we ran a series of 17 Friedman's tests, which is a non-parametric version of the repeated-measures ANOVAs, one per variable. This allowed us to test whether software engineers' well-being increased, decreased, or remained the same. 
Additional to the common descriptive (means and standard deviations) and inferential statistics ($\chi$-value and $p$-value\footnote{The $\chi$-value is a test-statistic that increases with larger mean-differences, lower within-group variability, or larger sample size. It is, for fixed sample size, inversely related to the p-value, which is used to determine whether our findings are statistically significant.}), we report as an effect size how many participants report a higher, lower, or equal level of any variable at Wave 6 compared to Wave 1. 
To correct the false-positive rate, which was necessary given the number of 17 tests, we applied the $M_{eff}$-correction method~\cite{derringer2018simple} which we resulted in a $\alpha$-level to .0037. We therefore only consider findings to be significant if $p < .004$. However, since there is, to the best of our knowledge, no single best way to correct for multiple comparison, we acknowledge that other researchers might prefer a more conservative or liberal threshold. Therefore, we report the exact $p$-values, allowing researchers to select different thresholds. 
Subsequently, we ran a series of 17 (dependent variables) $\times$ 7 (moderators) = 119 linear mixed-effects models to test whether past working from home experience, living alone (yes vs. no), income, gender (women vs. men), age, preference for working from home and having changed jobs predicted to change over time in any of the 17 dependent variables. Given the large number of tests, we set our $\alpha$-level to .0005. 

\subsubsection{Predictors of working from home}
To test which variables were associated with working from home, we ran a series of correlations that included the developers' preference for working from home, Changed jobs since March 2020, job satisfaction, solidarity, resilience, and intention to change jobs during our last data collection phase (i.e., wave 6 only). All 17 of the focal variables were measured in wave 6, as well as age and gender. Given the 24 tests, we again corrected our $\alpha$-level using the $M_{eff}$-correction method, this time to .0026. Additionally, this approach allowed us to explore predictors of intentions to change jobs and job satisfaction. 

We then tested whether the association of preference for working remotely with the intention to change jobs is moderated by company policy; we ran a moderated regression using the R-package interactions~\cite{interactions}. 

\subsubsection{Between-group comparisons}

Additionally, we compared people living in the United Kingdom and the USA (these were the two countries from which relatively most of our participants came) across all 17 variables and all 6 waves, resulting in $6 \times 17 = 102$ between-subject t-tests. Using again the $M_{eff}$-correction method, we adjusted our $alpha-$threshold to .0006. To address recent calls to report effect sizes that display similarities to avoid a one-sided focus on potentially small differences~\cite{hanel2019new}, we also report the effect size Percentages of Common Responses (PCR) alongside the more common effect size Cohen's $d$. PCR is a measure of overlap between two groups (e.g., British and US-American people) and ranges from 0 (no overlap/similarities) to 100 (both groups overlap perfectly). 
Finally, building on previous research that compared female and male software engineers~\cite{russo2020gender}. However, because in waves 5 and 6 few women participated (Table~\ref{tab:demographicsoverview}), we only compared women and men in waves $1--4$. Applying the $M_{eff}$-correction method, we adjusted our $alpha-$threshold to .0009.

\subsubsection{Thematic Analysis}

Our research provided highly relevant measurements to screen the tendency of the lockdown among developers. 
However, from the current design, we could not grasp more nuanced phenomena, typically emerging through a qualitative investigation.
Therefore, we performed a reflective thematic analysis with an inductive/deductive approach \cite{braun2006using,braun2019reflecting,braun2021one}.
This involved becoming familiar with the data through reading and re-reading transcripts, coding the data, developing and refining themes, and writing up the analysis. Throughout the process, the authors engaged in a consensus-based reflective and iterative process, constantly reviewing and refining the analysis to gain a deeper understanding of the data. 
This approach allowed for the emergence of themes or insights that may not have been apparent initially, leading to a more comprehensive understanding of the research topic.

The ultimate goal was to understand the initial thoughts of software engineering about their professional future in a post-pandemic context.

Consequently, during the last wave, when all surveyed countries were out of lockdowns, we included a reflective question about the future of hybrid work. In particular, we asked ``\textit{Many people expect that Hybrid Work is going to be the norm in the software industry. What are your thoughts on this?}''
In total, 103 informants provided a statement to our question.
Data were then deductively categorized as positive, neutral, or negative sentiments toward hybrid work. 

Afterward, we induced the individual codes into themes to grasp insights into personal beliefs of hybrid work.

\section{Results}
\label{sec:results}
In this section, we report the results of our analyses.
Details of the performed tests are in the online supplementary materials.

\subsection{Correlational analyses}
In the first step, we tested for construct validity by correlating all 17 variables with each other separately for each data collection wave. Results of the zero-order Pearson and Spearman correlations were as expected across all waves. For example, well-being correlated negatively with stress, loneliness, and boredom and positively with the need for autonomy, competence, and relatedness, which is in line with the literature~\cite{diener_beyond_2009,miller2011loneliness,russo2020predictors}.

\subsection{Changes along COVID-19 Pandemic}
\begin{figure}[h]
\centering
\includegraphics[width=1\linewidth]{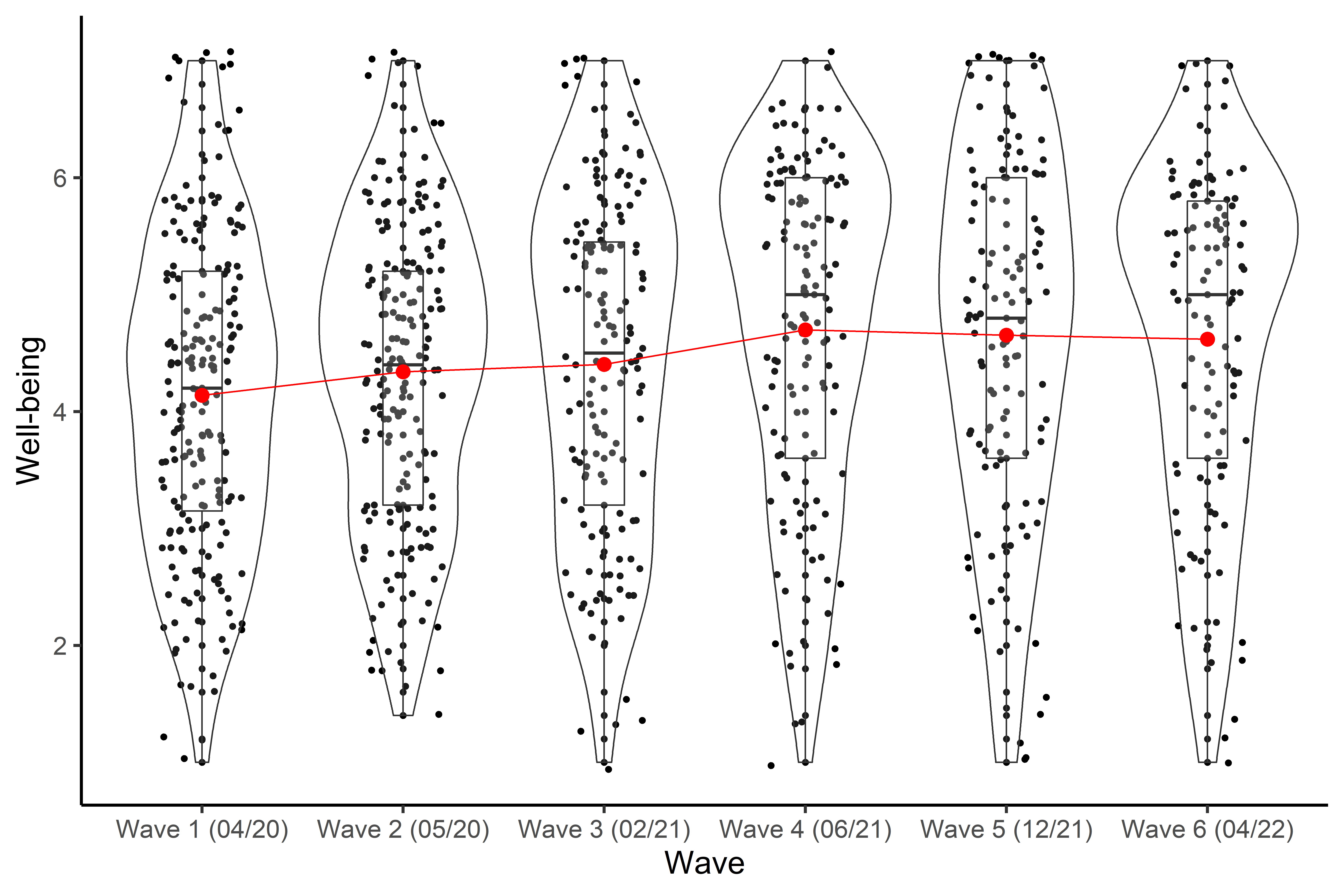}
\caption{Well-being across time. The red line displays the trend over time, whereas the box at each time point shows the range in which the middle 50\% of the data falls. Responses were given on a 7-point scale ranging from 1 to 7.}
\label{fig:wb}
\end{figure}

\begin{figure}[h]
\centering
\includegraphics[width=1\linewidth]{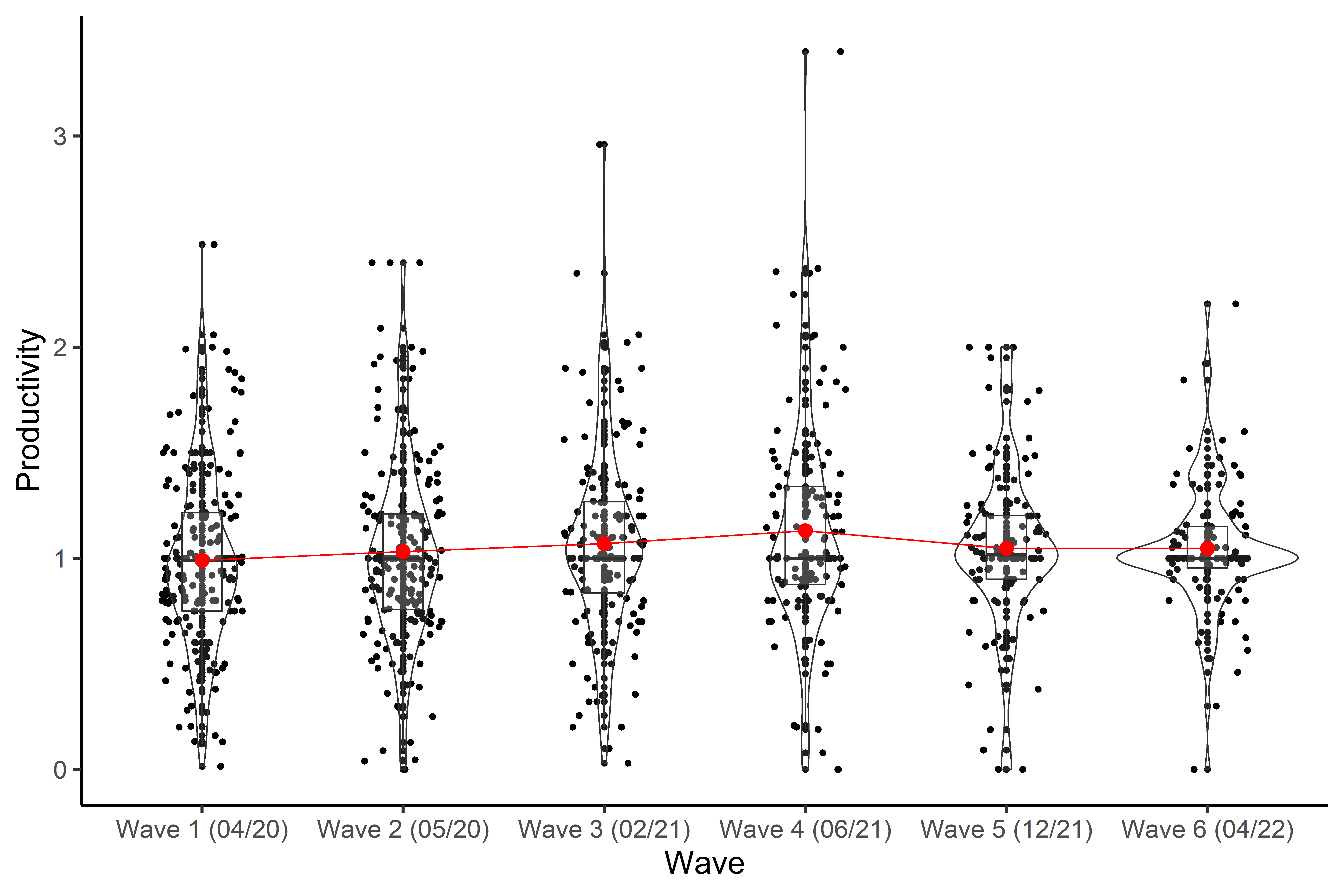} 
\caption{Productivity across time. The red line displays the trend over time, whereas the box at each time point shows the range in which the middle 50\% of the data falls. Here, a productivity score of one indicates that productivity has not changed compared to pre-pandemic levels. Scores $>$ 1 indicate that productivity increased and scores $<$ 1 indicate that productivity decreased. The figure highlights an increase in productivity across the presented timepoints.}
\label{fig:pr}
\end{figure}

The results of the 17 Friedman's test are displayed in Table~\ref{tab:ANOVAs}. Four of the tests were significant at $\alpha = .004$. Well-being (Fig~\ref{fig:wb}) increased, whereas emotional loneliness decreased (Fig~\ref{fig:el}). Behavioral disengagement decreased from wave 2 onwards (Fig~\ref{fig:bd}). Interestingly, the quality of social contacts (Fig~\ref{fig:sc}) showed a wave-form: It was lower during winter and early spring waves (i.e., April 2020, February 2021, and December 2021) than in later spring (i.e., May 2020, June 2021, and April 2022) when lockdowns were eased (i.e., in-person contacts were more accessible). 
For well-being, for example, 62 developers reported higher levels at Wave 6 than at Wave 1, 28 lower levels, and 11 an equal amount of well-being (cf. Tab.~\ref{tab:ANOVAs}). Interestingly, well-being increased between times 1 and 4, but then stayed relatively steady (Fig~\ref{fig:wb}). In contrast, productivity remained stable over time (Fig~\ref{fig:pr}).\footnote{Note that the Friedman's test necessarily only included the 73 participants that took part in all six waves whereas descriptive statistics reported in Table~\ref{tab:ANOVAs} and Figures~\ref{fig:wb} to ~\ref{fig:sc} all participants in each wave (e.g., 192 in wave 1) to improve prevision of the statistics we report in Appendix~\ref{sec:appendixC}.}

In the next step, we ran 17 (dependent variables) $\times$ 7 (moderators) = 119 linear mixed-effects models to test whether past working from home experience, living alone (yes vs. no), income, gender (women vs. men), age, preference for working from home and having changed jobs predicted to change over time in any of the 17 dependent variables. We also included dependent variables for which we found no main effect (i.e., no change over time) because it is possible that, for example, productivity increased for those who were living alone but decreased for those who were not. However, none of the 119 interaction terms reached statistical significance at $\alpha = .001$, all $ps > .005$.

\begin{table}[]
\caption{Friedman tests to test for changes over time for all 17 variables, significant variables at \textit{p} $\leq 0.004$ highlighted. $M_n$ represents the mean value of each wave and $SD_n$ its standard deviation. Rows with a highlighted background colour indicate a significant variable.}
\resizebox{\textwidth}{!}{
\begin{tabular}{@{}llllllllllllllllll@{}}
\toprule
\textbf{Variable}        & \textbf{M1} & \textbf{SD1} & \textbf{M2} & \textbf{SD2} & \textbf{M3} & \textbf{SD3} & \textbf{M4} & \textbf{SD4} & \textbf{M5} & \textbf{SD5} & \textbf{M6} & \textbf{SD6} & \textbf{$\chi$-value} & \textbf{p-value}  & \textbf{Greater} & \textbf{Lower} & \textbf{Equal} \\ 
\midrule
\rowcolor{g} Well-being               & 4.14        & 1.37         & 4.34        & 1.29         & 4.4         & 1.45         & 4.7         & 1.45         & 4.65        & 1.57         & 4.62        & 1.47         & 25.60             & $<.001$            & 62               & 28             & 11             \\
Productivity             & 0.99        & 0.42         & 1.03        & 0.44         & 1.07        & 0.44         & 1.13        & 0.51         & 1.05        & 0.39         & 1.05        & 0.31         & 4.24            & .5154           & 52               & 36             & 10             \\
Boredom                  & 2.94        & 1.14         & 2.93        & 1.16         & 2.83        & 1.27         & 2.77        & 1.18         & 2.77        & 1.26         & 2.69        & 1.18         & 5.50 & .3576           & 32               & 61             & 7              \\
\rowcolor{g} Behavioral disengagement & 1.8         & 0.94         & 2.06        & 1.03         & 1.88        & 1.11         & 1.84        & 1.07         & 1.91        & 1.15         & 1.74        & 1.05         & 18.50   & .0024        & 25               & 30             & 45             \\
Self-blame               & 1.81        & 0.99         & 1.88        & 1.01         & 2.28        & 1.29         & 2.25        & 1.26         & 2.31        & 1.26         & 2.1         & 1.2          & 8.83  & .1159               & 40               & 27             & 33             \\
Distractions at home     & 2.47        & 0.93         & 2.44        & 0.9          & 2.41        & 0.96         & 2.38        & 0.92         & 2.35        & 0.91         & 2.08        & 0.8          & 9.69 & .0847       & 23               & 49             & 27             \\
Generalized anxiety      & 2.25        & 1            & 2.17        & 1.01         & 2.2         & 1.07         & 2.1         & 1.06         & 2.01        & 0.97         & 1.97        & 0.96         & 16.43  & .0057              & 28               & 58             & 13             \\
\rowcolor{g} Emotional loneliness     & 2.11        & 0.9          & 2.01        & 0.87         & 2.1         & 0.91         & 1.88        & 0.9          & 1.86        & 0.82         & 1.67        & 0.75         &  31.01  &   $<.001$           & 20               & 58             & 21             \\
Social loneliness        & 2.64        & 1            & 2.56        & 1.02         & 2.79        & 1.08         & 2.73        & 1.04         & 2.83        & 1.01         & 2.55        & 1.14         & 14.74    & .0115       & 30               & 47             & 22             \\
Need for relatedness     & 3.5         & 0.83         & 3.56        & 0.8          & 3.48        & 0.84         & 3.59        & 0.82         & 3.51        & 0.77         & 3.54        & 0.82         & 6.25 & .2830           & 47               & 46             & 7              \\
Need for competence      & 3.57        & 0.74         & 3.58        & 0.73         & 3.62        & 0.76         & 3.67        & 0.74         & 3.63        & 0.71         & 3.76        & 0.75         & 6.32   & .2762        & 56               & 34             & 10             \\
Need for autonomy        & 3.48        & 0.69         & 3.51        & 0.73         & 3.42        & 0.77         & 3.51        & 0.77         & 3.44        & 0.76         & 3.46        & 0.72         & 4.07 & .5397            & 52               & 34             & 14             \\
\rowcolor{g} Social contacts          & 4.11        & 1.09         & 4.31        & 1.08         & 4.07        & 1.12         & 4.26        & 1.13         & 4.11        & 1.12         & 4.36        & 1.11         & 17.90 & .0031           & 56               & 28             & 15             \\
Communication            & 4.53        & 1            & 4.29        & 1.19         & 4.44        & 1.21         & 4.38        & 1.2          & 4.36        & 1.17         & 4.46        & 1.2          & 2.28 &   .8086         & 39               & 38             & 18             \\
Stress                   & 2.5         & 0.81         & 2.52        & 0.8          & 2.52        & 0.88         & 2.44        & 0.85         & 2.49        & 0.81         & 2.44        & 0.86         & 7.77 & .1691           & 36               & 47             & 17             \\
Daily routines           & 4.68        & 1.56         & 4.72        & 1.53         & 4.83        & 1.58         & 4.82        & 1.58         & 4.84        & 1.48         & 5.07        & 1.39         & 3.15    &  .6762     & 48               & 38             & 14             \\
Extraversion             & 3.45        & 0.79         & 3.46        & 0.78         & 3.47        & 0.8          & 3.46        & 0.71         & 3.45        & 0.78         & 3.44        & 0.8          & 2.51     & .7748      & 41               & 34             & 25             \\ 
\bottomrule
\end{tabular}
}
\label{tab:ANOVAs}
\end{table}

\subsection{Predictors of working from home}
Only one of the variables was associated with the preference for working from home at $\alpha = .003$, the perceived preference of the company $r(76) = .52, p < .001$, indicating that developers select a company that matches their preference regarding WFH or adjust to the company's preference. WFH was not associated with any other variables (Table \ref{tab:corrwfh}). This suggests that people who prefer WFH have not and do not intend to change jobs more often than those who prefer WFH less. Also, preference for WFH was not associated with satisfaction with the company, solidarity with the company, or resilience. \par

We found a range of other meaningful associations among the variables we only measured in wave 6. First, having changed jobs between March 2020 and April, 2022 predicted the intention to change jobs again. Further, general job satisfaction was positively associated with resilience, well-being, needs, social contact, communication, daily routines, and extraversion, as well as negatively with boredom, behavioral disengagement, self-blame, distractions, generalized anxiety, emotional and social loneliness, and stress. Solidarity with the company was positively associated with the same variables as job satisfaction. Resilience was positively associated with well-being, needs, quality of social contacts, communication, and extraversion, as well as negatively with boredom, behavioral disengagement, self-blame, distraction, generalized anxiety, emotional and social loneliness, and stress. Finally, the intention to change jobs was positively associated with boredom and negatively with autonomy.

\begin{table}[]
\caption{Correlations of preference for working from home. \\ \textit{Note}: Changed jobs since 03/2020 (0: N, 1: Y). Company preference for working in the office full time (0) vs working remotely full time (100). *$p < .05$, **$p < .01$, ***$p < .001$.}
\resizebox{\textwidth}{!}{
\begin{tabular}{@{}lllllll@{}}
\toprule
\textbf{} & \textbf{WFH} & \textbf{Changed jobs since 03/2020} & \textbf{Job satisfaction} & \textbf{Solidarity to the company} & \textbf{Resilience} & \textbf{Intention} \\ \midrule
Working from home          &        &         &         &         &         \\
Changed jobs               & -.05     &        &         &         &         &         \\
Job Satisfaction           & -.04     & .02    &         &         &         &         \\
Solidarity to Company      & -.25*    & -.16   & .3**    &         &         &         \\
Resilience                 & -.11     & 0      & .23*    & .37***  &         &         \\
Intention to change jobs   & .07      & .35*** & -.25*   & -.47*** & -.16    &         \\
Well-being                 & -.14     & -.02   & .36***  & .32**   & .42***  & -.26**  \\
Productivity               & -.03     & -.23*  & .07     & -.16    & -.14    & -.15    \\
Boredom                    & .13      & .14    & -.35*** & -.42*** & -.57*** & .37***  \\
Behavioral disengagement   & .14      & .1     & -.18    & -.3**   & -.51*** & .14     \\
Self-blame                 & .13      & -.01   & -.2     & -.26**  & -.48*** & .15     \\
Distraction                & -.07     & .16    & -.12    & -.33*** & -.35*** & .23*    \\
Generalized anxiety        & .16      & .12    & -.14    & -.3**   & -.54*** & .19     \\
Emotional loneliness       & .11      & .09    & -.21*   & -.3**   & -.37*** & .24*    \\
Social loneliness          & .17      & .07    & -.49*** & -.39*** & -.37*** & .3**    \\
Relatedness                & -.22*    & -.02   & .21*    & .34***  & .42***  & -.13    \\
Competence                 & -.12     & -.11   & .09     & .37***  & .47***  & -.29**  \\
Autonomy                   & -.14     & -.06   & .38***  & .55***  & .47***  & -.37*** \\
Social contact             & -.15     & -.01   & .39***  & .39***  & .39***  & -.28**  \\
Communication              & -.19     & .01    & .31**   & .53***  & .48***  & -.3**   \\
Stress                     & .08      & 0      & -.33*** & -.32**  & -.61*** & .28**   \\
Daily routines             & -.26**   & -.11   & .24*    & .24*    & .24*    & -.21*   \\
Extraversion               & -.16     & .13    & .22*    & .37***  & .45***  & .02     \\
Age                        & -.13     & -.17   & -.08    & .03     & .08     & -.24*   \\
Gender (1: M, 2: F)        & .08      & .07    & .01     & -.11    & -.16    & .18     \\ \bottomrule
\end{tabular}
}
\label{tab:corrwfh}
\end{table}

To test which variables predicted intention to change jobs, we ran a multiple regression analysis with all variables measured only in wave 6 as independent variables since they are most directly relevant. We also included autonomy and boredom because they were most strongly associated with the intention to change jobs, age, and gender. Results showed that only low job satisfaction and having changed jobs predicted the intention to change jobs (Table~\ref{tab:i2j}). It is worth noting that the majority of developers reported a low intention to change jobs (median = 2.50 on a 1-7 scale, see section~\ref{sec:measurements}).


\begin{table}[]
\centering
\caption{Regression coefficients of variables predicting intention to change jobs}
\begin{tabular}{@{}lllll@{}}
\toprule
                                 & \textbf{Estimate} & \textbf{Std. Error} & \textbf{t-value} & \textbf{p-value}     \\ \midrule
(Intercept)                      & 8.66     & 2.30       & 3.76    & 0.0004  *** \\
Working from home                & 0.00     & 0.01       & -0.20   & 0.8452      \\
Solidarity                       & -0.19    & 0.14       & -1.35   & 0.1833      \\
Job Satisfaction W6              & -1.12    & 0.32       & -3.46   & 0.0010  *** \\
Commuting                        & 0.01     & 0.13       & 0.11    & 0.9163      \\
Resilience                       & 0.25     & 0.26       & 0.95    & 0.3459      \\
Changed jobs in the past         & 1.20     & 0.38       & 3.16    & 0.0024  **  \\
Company preference regarding WFH & 0.00     & 0.01       & -0.58   & 0.5618      \\
Age W6                           & -0.02    & 0.02       & -1.00   & 0.3199      \\
Gender W6                        & 0.13     & 0.35       & 0.36    & 0.7185      \\
Boredom W6                       & -0.10    & 0.21       & -0.49   & 0.6290      \\
Autonomy W6                      & -0.23    & 0.36       & -0.64   & 0.5254      \\ \bottomrule
\end{tabular}
\label{tab:i2j}
\end{table}

\begin{table}[]
\centering
\caption{Regression coefficients of variables predicting job satisfaction}
\small
\begin{tabular}{@{}lllll@{}}
\toprule
                            & \textbf{Estimate} & \textbf{Std. Error} & \textbf{t value} & \textbf{p-value} \\ \midrule
(Intercept)                 & 2.55              & 1.40                & 1.82             & 0.0744           \\ \midrule
Working from home           & 0.00              & 0.00                & -1.22            & 0.2295           \\
Solidarity                  & 0.18              & 0.05                & 3.84             & 0.0003           \\
Commuting                   & 0.07              & 0.05                & 1.43             & 0.1595           \\
Resilience                  & 0.01              & 0.10                & 0.08             & 0.9397           \\
Changed jobs                & -0.10             & 0.13                & -0.78            & 0.4391           \\
Company  preference         & 0.00              & 0.00                & 0.53             & 0.5985           \\
Age W6                      & 0.00              & 0.01                & 0.38             & 0.7032           \\
Gender W6                   & -0.05             & 0.13                & -0.35            & 0.7253           \\
Well-being W6               & 0.13              & 0.07                & 1.83             & 0.0733           \\
Boredom W6                  & -0.12             & 0.08                & -1.50            & 0.1408           \\
Behavioral disengagement W6 & 0.13              & 0.08                & 1.56             & 0.1249           \\
Self blame W6               & -0.01             & 0.07                & -0.20            & 0.8444           \\
Distractions W6             & -0.17             & 0.08                & -2.12            & 0.0388           \\
Generalized anxiety W6      & -0.01             & 0.10                & -0.05            & 0.9588           \\
Emotional loneliness W6     & -0.11             & 0.12                & -0.89            & 0.3784           \\
Social loneliness W6        & -0.02             & 0.10                & -0.24            & 0.8131           \\
Relatedness W6              & -0.36             & 0.13                & -2.73            & 0.0086           \\
Competence W6               & -0.15             & 0.13                & -1.15            & 0.2556           \\
Autonomy W6                 & -0.07             & 0.15                & -0.46            & 0.6463           \\
Social contacts W6          & 0.11              & 0.12                & 0.92             & 0.3646           \\
Communication W6            & 0.31              & 0.07                & 4.15             & 0.0001           \\
Stress W6                   & 0.03              & 0.13                & 0.20             & 0.8435           \\
Daily routines W6           & 0.05              & 0.05                & 1.01             & 0.3164           \\
Extraversion W6             & 0.08              & 0.08                & 0.89             & 0.3756           \\ \bottomrule
\end{tabular}
\label{tab:js}
\end{table}

To test which variables predict job satisfaction, we again included the variables measured at wave 6 alongside the most strongly correlated with job satisfaction. Of the 14 predictors, only solidarity, $B = .18, SE = .05, p < .001$, distractions at home, $B = -.17, SE = .08, p = .04$, need for relatedness, $B = -.35, SE = .14, p = .01$, and communication with colleagues, $B = .31, SE = .08, p < .001$ were associated with job satisfaction (Table~\ref{tab:js}).  


We then tested whether the association of preference for working remotely with the intention to change jobs is moderated by company policy for working remotely vs. returning to the office full-time. The interaction was significant, $B = -.0004, SE = .0002, p = .043$. When the company preferred employees to work from the office, there was a positive association between their own preference for working remotely and intention to leave jobs, $B = .02, SE = .01, p = .05$. When the company's preference was more flexible towards working remotely, this association was non-significant, $B = -.01, SE = .01, p = .47$. This suggests that when the company wants their employees to return to the office but prefers working remotely, this can increase the intention to change jobs.

\subsection{Between-group comparisons}

The results of comparing people living in the United Kingdom and the USA in Table \ref{tab:ukus} (these were the two countries from which most of our participants came) across all 17 variables and all 6-time points. At Wave 1, our sample consisted of 63 people living in the UK and 52 in the USA. At Wave 4, 39 people living in the UK and 30 in the USA remained in the sample. None of the 102 between-country comparisons reached statistical significance at $\alpha = .0006$, all $ps > .03$. Similarities between groups were large, mean PCR = 94.46, $range = 77.30 - 99.92$.
Similarly, none of the gender comparison was significant at  $\alpha = .0009$, all $ps > .04$ (Table \ref{table_gender}). Similarities between women and men were large, mean PCR = 91.46, $range = 75.20 - 99.96$.

\begin{table*}[]
\caption{Comparisons between developers based in the United Kingdom and United States of America}
\label{tab:ukus}
\resizebox{\textwidth}{!}{%
\begin{tabular}{@{}lllllllllllllllll@{}}
\toprule
                         & \textbf{Wave 1} &                &               &                &                  &                  &                    &              & \textbf{Wave 2} &                &               &                &                  &                  &                    &              \\
                         & \textbf{UK M}   & \textbf{UK SD} & \textbf{US M} & \textbf{US SD} & \textbf{t-value} & \textbf{p-value} & \textbf{Cohen's d} & \textbf{PCR} & \textbf{UK M}   & \textbf{UK SD} & \textbf{US M} & \textbf{US SD} & \textbf{t-value} & \textbf{p-value} & \textbf{Cohen's d} & \textbf{PCR} \\ \midrule
Well-being               & 4.248           & 1.302          & 4.288         & 1.448          & -0.158           & 0.8752           & -0.03              & 98.803       & 4.294           & 1.22           & 4.392         & 1.461          & -0.381           & 0.7039           & -0.074             & 97.049       \\
Productivity             & 1.018           & 0.453          & 0.936         & 0.385          & 1.047            & 0.2975           & 0.193              & 92.312       & 0.977           & 0.414          & 1.076         & 0.472          & -1.162           & 0.248            & -0.225             & 91.043       \\
Boredom                  & 2.857           & 1.072          & 2.889         & 1.194          & -0.151           & 0.8802           & -0.029             & 98.843       & 2.96            & 1.159          & 2.74          & 1.166          & 0.994            & 0.3226           & 0.189              & 92.471       \\
Behavioral disengagement & 1.865           & 0.885          & 1.683         & 0.852          & 1.123            & 0.2641           & 0.21               & 91.638       & 2.089           & 0.952          & 1.91          & 1.024          & 0.948            & 0.3456           & 0.182              & 92.749       \\
Self blame               & 1.786           & 0.932          & 1.74          & 1.059          & 0.241            & 0.81             & 0.046              & 98.165       & 1.944           & 1.025          & 1.68          & 0.896          & 1.451            & 0.1498           & 0.272              & 89.182       \\
Relatedness              & 3.521           & 0.772          & 3.545         & 0.827          & -0.158           & 0.875            & -0.03              & 98.803       & 2.54            & 0.816          & 2.5           & 0.985          & 0.232            & 0.8168           & 0.045              & 98.205       \\
Competence               & 3.569           & 0.734          & 3.593         & 0.873          & -0.159           & 0.8742           & -0.03              & 98.803       & 2.219           & 0.965          & 1.989         & 0.962          & 1.257            & 0.2114           & 0.239              & 90.488       \\
Autonomy                 & 3.503           & 0.7            & 3.516         & 0.754          & -0.098           & 0.9223           & -0.018             & 99.282       & 2.059           & 0.938          & 1.887         & 0.796          & 1.052            & 0.2949           & 0.197              & 92.154       \\
Communication            & 4.472           & 1.031          & 4.593         & 1              & -0.624           & 0.5341           & -0.119             & 95.255       & 2.527           & 0.942          & 2.493         & 1.135          & 0.168            & 0.8673           & 0.032              & 98.723       \\
Stress                   & 2.528           & 0.713          & 2.312         & 0.871          & 1.43             & 0.156            & 0.273              & 89.143       & 3.538           & 0.711          & 3.593         & 0.847          & -0.372           & 0.7111           & -0.072             & 97.128       \\
Daily routines           & 4.889           & 1.409          & 4.474         & 1.738          & 1.385            & 0.1693           & 0.265              & 89.459       & 3.605           & 0.67           & 3.65          & 0.815          & -0.315           & 0.7534           & -0.061             & 97.567       \\
Extraversion             & 3.552           & 0.728          & 3.486         & 0.799          & 0.459            & 0.6473           & 0.087              & 96.53        & 3.565           & 0.751          & 3.443         & 0.818          & 0.808            & 0.421            & 0.155              & 93.823       \\
Distractions             & 2.532           & 0.92           & 2.385         & 1.018          & 0.806            & 0.4222           & 0.152              & 93.942       & 4.274           & 1.12           & 4.327         & 1.189          & -0.238           & 0.8121           & -0.046             & 98.165       \\
Generalized anxiety      & 2.265           & 0.942          & 2.134         & 1.075          & 0.689            & 0.4926           & 0.131              & 94.778       & 4.383           & 1.237          & 4.245         & 1.263          & 0.573            & 0.5678           & 0.11               & 95.614       \\
Emotional loneliness     & 2.048           & 0.956          & 2.038         & 0.802          & 0.056            & 0.9555           & 0.01               & 99.601       & 2.573           & 0.69           & 2.34          & 0.89           & 1.516            & 0.133            & 0.296              & 88.234       \\
Social loneliness        & 2.619           & 0.912          & 2.583         & 1.074          & 0.19             & 0.8498           & 0.036              & 98.564       & 4.817           & 1.531          & 4.647         & 1.646          & 0.562            & 0.5752           & 0.108              & 95.694       \\
Social contacts          & 4.053           & 1.091          & 4.218         & 1.064          & -0.818           & 0.415            & -0.153             & 93.902       & 3.516           & 0.742          & 3.53          & 0.798          & -0.094           & 0.925            & -0.018             & 99.282       \\ \midrule
                         & \textbf{Wave 3} &                &               &                &                  &                  &                    &              & \textbf{Wave 4} &                &               &                &                  &                  &                    &              \\
                         & \textbf{UK M}   & \textbf{UK SD} & \textbf{US M} & \textbf{US SD} & \textbf{t-value} & \textbf{p-value} & \textbf{Cohen's d} & \textbf{PCR} & \textbf{UK M}   & \textbf{UK SD} & \textbf{US M} & \textbf{US SD} & \textbf{t-value} & \textbf{p-value} & \textbf{Cohen's d} & \textbf{PCR} \\ \midrule
Well-being               & 4.508           & 1.18           & 4.611         & 1.616          & -0.323           & 0.7481           & -0.074             & 97.049       & 4.751           & 1.325          & 4.733         & 1.554          & 0.051            & 0.9596           & 0.013              & 99.481       \\
Productivity             & 1.103           & 0.511          & 1.081         & 0.439          & 0.211            & 0.8331           & 0.046              & 98.165       & 1.103           & 0.511          & 1.081         & 0.439          & 0.211            & 0.8331           & 0.046              & 98.165       \\
Boredom                  & 2.865           & 1.135          & 2.413         & 1.148          & 1.792            & 0.0772           & 0.396              & 84.305       & 2.79            & 1.161          & 2.558         & 1.218          & 0.806            & 0.4234           & 0.195              & 92.233       \\
Behavioral disengagement & 1.875           & 1.137          & 1.722         & 1.168          & 0.6              & 0.5502           & 0.133              & 94.698       & 1.817           & 1.053          & 1.833         & 1.155          & -0.061           & 0.9517           & -0.015             & 99.402       \\
Self blame               & 2.271           & 1.12           & 1.889         & 1.321          & 1.398            & 0.1665           & 0.316              & 87.446       & 2.305           & 1.298          & 2.117         & 1.112          & 0.656            & 0.5141           & 0.154              & 93.862       \\
Distractions             & 2.354           & 0.844          & 2.25          & 0.914          & 0.534            & 0.595            & 0.119              & 95.255       & 2.329           & 0.826          & 2.367         & 0.964          & -0.171           & 0.8646           & -0.042             & 98.325       \\
Generalized anxiety      & 2.193           & 1.006          & 2.095         & 1.137          & 0.411            & 0.682            & 0.092              & 96.331       & 2.024           & 0.963          & 2.048         & 1.175          & -0.089           & 0.9297           & -0.022             & 99.122       \\
Emotional loneliness     & 2.083           & 0.86           & 2.028         & 0.977          & 0.271            & 0.787            & 0.061              & 97.567       & 1.959           & 0.895          & 1.744         & 0.87           & 1.016            & 0.3135           & 0.243              & 90.33        \\
Social loneliness        & 2.674           & 0.903          & 2.491         & 1.197          & 0.768            & 0.4456           & 0.176              & 92.988       & 2.553           & 0.89           & 2.722         & 1.135          & -0.679           & 0.5002           & -0.169             & 93.266       \\
Relatedness              & 3.479           & 0.764          & 3.718         & 0.799          & -1.379           & 0.1719           & -0.306             & 87.84        & 3.65            & 0.714          & 3.689         & 0.882          & -0.197           & 0.8449           & -0.049             & 98.045       \\
Competence               & 3.632           & 0.643          & 3.764         & 0.79           & -0.819           & 0.4158           & -0.186             & 92.59        & 3.65            & 0.694          & 3.672         & 0.839          & -0.116           & 0.9079           & -0.029             & 98.843       \\
Autonomy                 & 3.403           & 0.716          & 3.523         & 0.832          & -0.696           & 0.4888           & -0.157             & 93.743       & 3.614           & 0.706          & 3.567         & 0.711          & 0.277            & 0.7827           & 0.067              & 97.328       \\
Social contacts          & 4.014           & 1.034          & 4.38          & 1.042          & -1.597           & 0.1144           & -0.353             & 85.99        & 4.374           & 1.123          & 4.322         & 1.049          & 0.199            & 0.8426           & 0.047              & 98.125       \\
Communication            & 4.299           & 1.283          & 4.81          & 1.011          & -2.028           & 0.0459           & -0.434             & 82.821       & 4.275           & 1.322          & 4.444         & 0.996          & -0.612           & 0.5428           & -0.142             & 94.34        \\
Stress                   & 2.547           & 0.738          & 2.403         & 0.955          & 0.753            & 0.4545           & 0.172              & 93.147       & 2.409           & 0.784          & 2.35          & 0.829          & 0.301            & 0.7648           & 0.073              & 97.088       \\
Daily routines           & 4.944           & 1.6            & 5.009         & 1.552          & -0.187           & 0.8522           & -0.041             & 98.364       & 5.122           & 1.533          & 4.756         & 1.744          & 0.92             & 0.3615           & 0.226              & 91.003       \\
Extraversion             & 3.568           & 0.766          & 3.576         & 0.843          & -0.049           & 0.9614           & -0.011             & 99.561       & 3.591           & 0.72           & 3.517         & 0.707          & 0.437            & 0.6636           & 0.105              & 95.813       \\ \bottomrule
\end{tabular}%
}
\end{table*}
\begin{table*}[]
\caption{Comparisons between women and men. M = mean, SD = standard deviation.}
\label{tab:menwomen}
\resizebox{\textwidth}{!}{%
\begin{tabular}{@{}lllllllllllllllll@{}}
\toprule
 &  \textbf{Wave 1} &   &   &   &   &   &   &   &  \textbf{Wave 2} &   &   &   &   &   &   &   \\
 &  \textbf{Men M} &  \textbf{Men SD} &  \textbf{Women M} &  \textbf{Women SD} &  \textbf{t-value} &  \textbf{p-value} &  \textbf{Cohen's d} &  \textbf{PCR} &  \textbf{Men M} &  \textbf{Men SD} &  \textbf{Women M} &  \textbf{Women SD} &  \textbf{t-value} &  \textbf{p-value} &  \textbf{Cohen's d} &  \textbf{PCR} \\ \midrule
Well-being &  4.109 &  1.336 &  4.263 &  1.495 &  -0.581 &  0.5639 &  -0.113 &  95.494 &  4.388 &  1.258 &  4.151 &  1.407 &  0.933 &  0.3553 &  0.183 &  92.71 \\
Productivity &  1.008 &  0.416 &  0.917 &  0.43 &  1.18 &  0.243 &  0.218 &  91.32 &  1.029 &  0.433 &  1.043 &  0.452 &  -0.175 &  0.862 &  -0.033 &  98.684 \\
Boredom &  2.942 &  1.13 &  2.908 &  1.179 &  0.163 &  0.8713 &  0.03 &  98.803 &  2.923 &  1.122 &  2.943 &  1.309 &  -0.082 &  0.9353 &  -0.016 &  99.362 \\
Behavioral   disengagement &  1.799 &  0.935 &  1.829 &  0.953 &  -0.176 &  0.8611 &  -0.032 &  98.723 &  1.997 &  0.957 &  2.324 &  1.259 &  -1.479 &  0.1458 &  -0.32 &  87.288 \\
Self blame &  1.753 &  0.957 &  2.053 &  1.095 &  -1.546 &  0.1283 &  -0.304 &  87.919 &  1.83 &  0.955 &  2.081 &  1.211 &  -1.173 &  0.2465 &  -0.248 &  90.132 \\
Relatedness &  3.483 &  0.801 &  3.557 &  0.948 &  -0.446 &  0.6577 &  -0.089 &  96.451 &  2.459 &  0.869 &  2.378 &  1.003 &  0.449 &  0.655 &  0.09 &  96.411 \\
Competence &  3.566 &  0.704 &  3.596 &  0.862 &  -0.202 &  0.8407 &  -0.041 &  98.364 &  2.098 &  0.963 &  2.475 &  1.144 &  -1.844 &  0.0712 &  -0.376 &  85.088 \\
Autonomy &  3.476 &  0.668 &  3.509 &  0.771 &  -0.239 &  0.8119 &  -0.047 &  98.125 &  1.975 &  0.837 &  2.135 &  0.998 &  -0.899 &  0.373 &  -0.184 &  92.67 \\
Communication &  4.511 &  1.004 &  4.623 &  0.972 &  -0.625 &  0.5343 &  -0.112 &  95.534 &  2.56 &  0.985 &  2.577 &  1.151 &  -0.08 &  0.9365 &  -0.016 &  99.362 \\
Stress &  2.468 &  0.744 &  2.638 &  1.025 &  -0.966 &  0.3391 &  -0.212 &  91.558 &  3.56 &  0.748 &  3.554 &  1.004 &  0.034 &  0.9728 &  0.008 &  99.681 \\
Daily routines &  4.758 &  1.469 &  4.368 &  1.877 &  1.191 &  0.2394 &  0.25 &  90.052 &  3.605 &  0.692 &  3.491 &  0.874 &  0.74 &  0.4628 &  0.156 &  93.783 \\
Extraversion &  3.401 &  0.786 &  3.638 &  0.766 &  -1.7 &  0.0944 &  -0.303 &  87.958 &  3.526 &  0.698 &  3.45 &  0.863 &  0.494 &  0.6236 &  0.103 &  95.893 \\
Distractions &  2.481 &  0.884 &  2.408 &  1.126 &  0.37 &  0.7127 &  0.078 &  96.889 &  4.281 &  1.055 &  4.432 &  1.165 &  -0.719 &  0.4754 &  -0.14 &  94.419 \\
Generalized   anxiety &  2.123 &  0.916 &  2.738 &  1.175 &  -3.007 &  0.0042 &  -0.632 &  75.2 &  4.293 &  1.118 &  4.288 &  1.434 &  0.02 &  0.9839 &  0.004 &  99.84 \\
Emotional loneliness &  2.022 &  0.848 &  2.474 &  1.033 &  -2.498 &  0.0158 &  -0.51 &  79.872 &  2.485 &  0.748 &  2.662 &  0.965 &  -1.042 &  0.3025 &  -0.223 &  91.122 \\
Social   loneliness &  2.667 &  0.969 &  2.535 &  1.146 &  0.653 &  0.5169 &  0.131 &  94.778 &  4.902 &  1.439 &  3.982 &  1.689 &  3.049 &  0.0037 &  0.617 &  75.77 \\
Social contacts &  4.032 &  1.068 &  4.421 &  1.149 &  -1.893 &  0.0637 &  -0.358 &  85.794 &  3.405 &  0.775 &  3.662 &  0.769 &  -1.817 &  0.0745 &  -0.333 &  86.776 \\ \midrule
 &  \textbf{Wave 3} &   &   &   &   &   &   &   &  \textbf{Wave 4} &   &   &   &   &   &   &   \\
 &  \textbf{Men M} &  \textbf{Men SD} &  \textbf{Women M} &  \textbf{Women SD} &  \textbf{t-value} &  \textbf{p-value} &  \textbf{Cohen's d} &  \textbf{PCR} &  \textbf{Men M} &  \textbf{Men SD} &  \textbf{Women M} &  \textbf{Women SD} &  \textbf{t-value} &  \textbf{p-value} &  \textbf{Cohen's d} &  \textbf{PCR} \\ \midrule
Well-being &  4.456 &  1.418 &  4.207 &  1.576 &  0.787 &  0.4356 &  0.172 &  93.147 &  4.718 &  1.479 &  4.63 &  1.341 &  0.294 &  0.7698 &  0.061 &  97.567 \\
Productivity &  1.057 &  0.413 &  1.116 &  0.555 &  -0.542 &  0.5912 &  -0.132 &  94.738 &  1.057 &  0.413 &  1.116 &  0.555 &  -0.542 &  0.5912 &  -0.132 &  94.738 \\
Boredom &  2.786 &  1.253 &  3 &  1.361 &  -0.778 &  0.4408 &  -0.168 &  93.306 &  2.774 &  1.163 &  2.741 &  1.243 &  0.126 &  0.9 &  0.029 &  98.843 \\
Behavioral disengagement &  1.798 &  1.057 &  2.183 &  1.263 &  -1.535 &  0.1327 &  -0.349 &  86.147 &  1.799 &  1.082 &  1.981 &  1.014 &  -0.815 &  0.4195 &  -0.171 &  93.186 \\
Self   blame &  2.167 &  1.208 &  2.7 &  1.495 &  -1.805 &  0.0787 &  -0.419 &  83.406 &  2.119 &  1.224 &  2.741 &  1.296 &  -2.232 &  0.0313 &  -0.502 &  80.181 \\
Distractions &  2.408 &  0.921 &  2.417 &  1.099 &  -0.04 &  0.9682 &  -0.009 &  99.641 &  2.381 &  0.946 &  2.352 &  0.83 &  0.159 &  0.8745 &  0.032 &  98.723 \\
Generalized   anxiety &  2.099 &  1.005 &  2.605 &  1.245 &  -2.055 &  0.0466 &  -0.478 &  81.111 &  2.012 &  1.027 &  2.397 &  1.133 &  -1.592 &  0.1194 &  -0.366 &  85.48 \\
Emotional loneliness &  2.05 &  0.883 &  2.311 &  1.017 &  -1.286 &  0.2055 &  -0.287 &  88.59 &  1.845 &  0.904 &  2.012 &  0.908 &  -0.846 &  0.4025 &  -0.185 &  92.63 \\
Social   loneliness &  2.789 &  1.071 &  2.789 &  1.153 &  0.003 &  0.998 &  0.001 &  99.96 &  2.708 &  1 &  2.827 &  1.178 &  -0.48 &  0.634 &  -0.115 &  95.415 \\
Relatedness &  3.554 &  0.793 &  3.2 &  0.952 &  1.873 &  0.0684 &  0.428 &  83.055 &  3.624 &  0.828 &  3.481 &  0.814 &  0.8 &  0.4283 &  0.172 &  93.147 \\
Competence &  3.664 &  0.716 &  3.444 &  0.892 &  1.245 &  0.2204 &  0.29 &  88.471 &  3.732 &  0.684 &  3.426 &  0.887 &  1.66 &  0.1057 &  0.418 &  83.445 \\
Autonomy &  3.488 &  0.745 &  3.144 &  0.807 &  2.11 &  0.0407 &  0.454 &  82.042 &  3.522 &  0.789 &  3.488 &  0.72 &  0.217 &  0.8295 &  0.045 &  98.205 \\
Social   contacts &  4.099 &  1.085 &  3.944 &  1.26 &  0.616 &  0.5411 &  0.138 &  94.499 &  4.323 &  1.09 &  4.012 &  1.259 &  1.166 &  0.2509 &  0.275 &  89.064 \\
Communication &  4.444 &  1.171 &  4.402 &  1.373 &  0.152 &  0.8803 &  0.035 &  98.604 &  4.372 &  1.199 &  4.407 &  1.221 &  -0.132 &  0.8956 &  -0.029 &  98.843 \\
Stress &  2.471 &  0.858 &  2.717 &  0.96 &  -1.272 &  0.2104 &  -0.279 &  88.906 &  2.392 &  0.885 &  2.63 &  0.708 &  -1.457 &  0.1513 &  -0.28 &  88.866 \\
Daily routines &  5.029 &  1.434 &  4.078 &  1.875 &  2.588 &  0.0136 &  0.62 &  75.656 &  5.003 &  1.446 &  4.148 &  1.884 &  2.186 &  0.0356 &  0.552 &  78.255 \\
Extraversion &  3.454 &  0.766 &  3.517 &  0.94 &  -0.337 &  0.7377 &  -0.078 &  96.889 &  3.443 &  0.727 &  3.528 &  0.663 &  -0.573 &  0.5692 &  -0.118 &  95.295 \\ \bottomrule
\end{tabular}
}
\end{table*}

\subsection{Thematic Analysis}

A vast majority of the surveyed informants reported a positive sentiment towards hybrid work in their future working lives (84.5\%).
They provided multiple reasons for this positive sentiment.
A common motivation highlighted by the participants is the observation that the software industry is particularly well suited for hybrid work:

\begin{displayquote}
It is an industry where this makes sense. I think this will end up becoming the norm also. [\#19]
\end{displayquote}

Others mention the possibility for a balance between more focused work at home and social interaction in the office:

\begin{displayquote}
I like the idea of hybrid work because certain tasks get done faster in person, and it is easier to network in person compared to virtually. [\#49]
\end{displayquote}

Some professionals expressed the believe that hybrid work is more convenient for them and allow a better work-life balance:

\begin{displayquote}
I enjoy hybrid work because I get to see and interact with coworkers [...] and spend more time with my family. [\#58]
\end{displayquote}

Management support is considered a key enabler of hybrid work. 
When organizations can involve developers and create a sense of belonging, informants are optimistic about the future outlook of hybrid work:

\begin{displayquote}
Totally agree. [...] There is a sense of letting people work from home but making sure they are still motivated and feel a sense of belonging to the company. [\#84]
\end{displayquote}

All in all, there are several reasons why software engineers seem so positive.
Companies might save resources on rentals and workspace management.
At the same time, it is much more convenient for employees and improves work-life balance.
Also, several respondents stated their belief that a hybrid work setting provides a tremendous productivity and well-being booster.
Although project management changes are not trivial, they are considered addressable, especially with genuine support from management:

\begin{displayquote}
I hope it will be because I prefer it. I feel there are benefits for companies as well, i.e., less demands on office space, no need to provide a computer, desk, and chair. I was often tired and listless due to commuting, and I think my company benefits from healthier and happier employees. In terms of programming, at home, I feel I can work uninterrupted but also take a walk, or make a drink, etc, if I need it to recharge my batteries. There are still challenges for companies, for example, in ensuring that the employees are on-task and productive. But this is just project management. My managers have been clear with me about deadlines and monitor my progress through regular meetings. It seems to work, I'm productive, and things get done. Also, at home, I get new ideas for how things can be improved, and I can look into them a bit and feed them back to my team. So you could argue that working from home makes me more creative and innovative. At work, it's hard to be creative and innovative when you are being interrupted all the time, though I do understand that certain personality types do thrive in those very conditions. [\#77]
\end{displayquote}

A negative sentiment towards hybrid work is shared only among 6.8\% of our informants.
In those cases, there is a lack of trust in their organizations to restructure the internal processes and culture to fit hybrid work practices:

\begin{displayquote}
Unlikely due to management culture. [\#62]
\end{displayquote}

Personal preferences do also play a role. 
Some are more acquainted with the office environment and perceive daily physical interaction with colleagues as pivotal for their productivity:

\begin{displayquote}
Personally, I prefer to be in an office environment. [...] It's far better, in my opinion, to work in an office environment with others, even though I'm an introvert and like spending my free time on my own. [\#95]
\end{displayquote}

Finally, some developers have a neutral sentiment towards hybrid work (8.7\%).
Some of our respondents point to their local culture as playing a role in the adoption of hybrid work practices. 
Those with a flat hierarchy are more likely successfully work hybrid:

\begin{displayquote}
I think it depends from country to country. More progressive countries will likely keep some level of remote work, while more conservative countries will probably want people back in offices. [\#2]
\end{displayquote}

Similarly, company culture also plays a significant role in the likelihood of adopting hybrid policies:

\begin{displayquote}
[...] There is also still this old school mentality of ``if you're not in the office at your desk, you're not working" that is still pretty prevalent, at least where I am, so I can see that hybrid might work as a compromise for those business leaders who may not want to allow their workforce to be fully remote. [\#8]
\end{displayquote}

The type of task or project software engineers have to perform can also be a variable where hybrid might be a suitable option:

\begin{displayquote}
It may be beneficial in some cases, depending on the type of project and, more importantly, the project phase (specs, coding, testing). [\#40]
\end{displayquote}

\begin{displayquote}
It depends on the role of the job. For managing hardware and servers, you do sometimes need to be there in person. For programming/coding and such like there is no need to actually be in an office, so this could be done 100\% of the time from any location. [\#78]
\end{displayquote}

To conclude, \textbf{there is a strong sentiment among developers that hybrid work will be the new norm in the tech industry}.
This sentiment is widely shared among the professional community, as similar investigations suggest~\cite{smite2023work,Smite2023workplace}. 
Several issues emerged from our informants.
First, \textbf{company culture and management support} is a clear enabler or inhibitor for a hybrid transition.
Second, \textbf{type of task} is considered a decisive factor when hybrid work agreements since some are not seen suited for it.
Third, how \textbf{face-to-face communication} opportunities will happen in organizations and the threat of jeopardized contacts is a widely shared concern. 
Although developers feel confident working remotely, they also believe synchronization and alignment opportunities are crucial to performing the assigned task.

\section{Discussion}
\label{sec:discussion}

Building on the collected evidence and the previous literature, we discuss the implications of our investigation for software professionals and organizations.
Furthermore, we explain the intrinsic limitations of this study and how we tried to cope with them.

First, however, readers should be aware that our findings are based on group-level inferences, which do not always generalize to the individual level. For example, the results of the Friedman's test inform us whether the \textit{average} of a variable changed over time, not whether all individuals changed in the same direction. As seen in Table \ref{tab:ANOVAs}, while the well-being of 62 developers increased between Waves 1 and 6, the well-being of 28 developers dropped. Thus, it is over twice as likely to find a developer whose well-being increased instead of dropped. 
Interestingly, the change over time was not always linear. For example, emotional loneliness went down between Wave 1 and Wave 2, then slightly up again at Wave 3, before declining until wave 6. This might be because many countries started to (announce plans to) open up again around the time when we collected the second wave. In contrast, the third wave was collected in February 2021: In the UK and USA, for example, in winter 2020/21, the deaths of many more people were associated with COVID-19 compared to spring 2020. During waves 4-6, more and more people got vaccinated, allowing more in-person interactions. We observed a similar pattern for the quality of social contacts.
Similarly, there are a range of situational factors or variables which we have not measured, such as the perceived severity of local lockdowns or loss of a loved one (e.g., because of COVID-19) that would likely have explained additional variance in developers' well-being and productivity~\cite{hu2022disruptive}. 
Nevertheless, we aimed to provide generalizable evidence with this longitudinal study. Our qualitative investigations (e.g.,~\cite{miller2021your,ford2020tale,butler2021challenges}) add to a nuanced understanding of individual phenomena.

Further, our results are overall in line with similar other longitudinal research. For example, well-being increased while anxiety decreased from early April onwards over a period of six weeks in a large sample of people living in Great Britain~\cite{o2021mental}. This pattern was mostly replicated in a French sample, even though well-being initially dropped at the onset of the lockdown before it went up again in the following weeks~\cite{pellerin2020psychological}. In a sample of teachers well-being kept dropping during the first 7 months of the pandemic, presumably because their stress-levels intensified more compared to other professions~\cite{kim2022my}. Thus, our findings are overall in line with similar studies that included people from the general public.
These and other studies should also be considered when drawing company guidelines since our recommendations will be partial.

\subsection{Implications}

Based on our results, we provide recommendations for the software engineering community (cf. also Table \ref{tab:summary}).

We found that developers' \textbf{well-being increased over time}. 
We have no pre-pandemic data, so we can not assess how the lockdown initially impacted software professionals. 
It could be that their well-being went down in the Spring of 2020 and is now bouncing back to pre-pandemic times. 
This reasoning would align with previous research showing that people's well-being usually bounces back after a significant negative event~\cite{oswald2008does} and research showing that work and family satisfaction declined between 2019 and April 2020~\cite{mohring2021covid}. 
While research from the start of the pandemic (i.e., Spring 2020) indicates that developers' well-being decreased initially~\cite{Ralph2020pandemic}, our findings provide a more positive outlook that developers' well-being bounced back.
Our findings also suggest that working from home does not negatively impact developers' well-being: otherwise, well-being would not have increased between waves 1 to 4. 
Our data show how well-being increased until June 2021 and then remained constant till the end of the pandemic (cf. Figure \ref{fig:wb}).
Furthermore, working from home was in wave 6 uncorrelated with well-being.
This indicates that software engineers learned to cope with the new enforced setting.
As a confirmation to that, more relaxed public health policies implemented from the Summer 2021 did not affect significantly developers' well-being.

\textbf{Productivity} remained constant during the pandemic.
Although we report a slight increase in productivity over the six data collection waves (as plotted in Figure \ref{fig:pr}), and more people reported an increase in productivity compared to those who reported a decrease (cf. Table \ref{tab:ANOVAs}), the mean differences are non-significant, indicating that the observed increase could very well be random and might not replicate.
Since measuring productivity is non-trivial, we followed a previous study example~\cite{russo2020predictors} by measuring productivity as a self-reported function compared to the pre-pandemic period.
Therefore, we conclude that software professionals' productivity level remained the same throughout the lockdown and compared to the pre-pandemic time.
This finding also contradicts previous research suggesting that the lockdown is detrimental to productivity~\cite{ralph2020empirical}, possibly because of differences in the research design (cross-sectional vs. longitudinal design) and operationalization of productivity (Ralph et al.~\cite{ralph2020empirical} used a different measure of productivity).
Additionally, we collected our sample approximately one month after Ralph et al.~\cite{ralph2020empirical} and predominantly from relatively underrepresented countries in the sample of Ralph et al., who recruited most of their participants from Germany, Russia, and Brazil, which might further explain why our findings do not align with those from Ralph et al. 
Our results substantiate our previous conclusion that a hybrid or full remote working environment would not \textit{per se} harm the productivity levels of developers.

Even though all typical welfare support (e.g., childcare, schools, and sports facilities) was closed, software engineers showed a high level of adaptation by keeping the same productivity levels and steadily increasing their well-being levels.
Consequently, in a post-pandemic working from home context, with all support facilities normally running, working from home is unlikely to impact developers' well-being negatively on average. This means, at the same time, that some developers might feel more lonely or experience other negative emotions when working from home, while others enjoy it more. This is likely to depend on various factors such as their quality of social contact, how happy they are to (not) see their colleagues, office set-ups at home, or difficulty of their commute. 
Qualitative findings support this argument, suggesting that working from home significantly improved work-life balance~\cite{ford2020tale}.
Similarly, a large-scale cross-sectional study observed that $89$\% of the surveyed professionals would like to continue to work remotely (especially in a hybrid fashion)~\cite{Walton2020NZadaptation}.
However, previous research regarding the impact of working from home on productivity is mixed. Some studies found that working from home is positively or unrelated to productivity~\cite{barrero2021working,deole2021home,russo2020predictors}, whereas other research found that working from home has some negative effects~\cite{gibbs2021work,kitagawa2021working,morikawa2020productivity}). Interestingly, the studies which found that working from home is negatively associated with productivity were conducted in India and Japan, whereas those which found positive effects in the UK and USA. This finding provides some support for the individualism-collectivism hypothesis~\cite{hofstede2001culture}: People in collectivist cultures might struggle more to work alone and prefer working in teams. However, this reasoning is speculative as the individualism-collectivism hypothesis has been increasingly challenged in recent years~\cite{takano2018comparing}.

Software professionals \textbf{felt less lonely and improved their social contacts}.
During the first lockdown in Spring 2020, many people had to reduce their social interactions abruptly~\cite{chan2021can}.
As a consequence, this increased the sense of loneliness and isolation.
Nevertheless, also, in this case, developers showed a high level of resilience.
Indeed, we report a significant decrease in emotional loneliness and an increase in the quality of social contact.
This is a possible indication that software engineers increasingly reached out to their social contacts when they felt lonely, thereby coping well with the challenging conditions of the pandemic.
Similarly, the quality of their relationships increased.
This is important because having a reliable social support network is an essential coping mechanism, especially in hard times and in moments of high stress~\cite{weinstein2011self,carver1989assessing}.

These findings are relevant for organizations planning to implement a hybrid or remote work policy.
Software engineers showed a high level of resilience when coping with unexpected events.
At the same time, their social network was of crucial support while working from home.
This insight is also supported by previous research, where communication was found to be a relevant predictor for developers' satisfaction during the lockdown~\cite{miller2021your}.
Consequently, a proactive company policy of employee inclusion would sustain their well-being levels.
This would require a particular effort from the middle management (because they are the direct company interface for each employee) to ensure that every team member can express themselves and maintain stimulating and nurturing relationships with their peers since even interacting with weak social ties (i.e., acquaintances) can improve people's well-being~\cite{sandstrom2014social}.


\textbf{We found no mean differences between the UK and USA}.
We found high levels of similarities in how software professionals were impacted in the USA and the UK.
This might be the result of the reliance of national health authorities on the World Health Organization, making lockdown measures fairly uniform between both countries.

\textbf{Preference for working from home} was only associated with developers' perceived preference of the company regarding their WFH policy. 
Based on previous research on value change~\cite{bardi2014value}, we derive that both self-selection as well as socialization effects play a role here. 
Consequently, the preference for working from home might depend on self-selection (developers only start to work at companies that match their own preference) or socialization effects (developers preference change while working at a company). 
Indeed, preference for working from home was not associated with satisfaction and solidarity with the company. This non-significant finding is in our view encouraging because it shows that people who prefer working from home are equally satisfied and in solidarity with the company as well as equally productive than developers who are less keen to work from home.
Interestingly, we found that when the company preferred employees to work from the office, developers who preferred working remotely were more likely to change jobs. This suggests that a mismatch between employees and employers expectations regarding remote work can increase the chance of resignation.

This point is tightly connected to \textbf{the Future of Work}.
Most our informants agree that it will be different from what it used to be before the pandemic.
How different it will be is a question that will be asked with time.
There is, however, a general consensus in the scholarly community that the future of work is going to look different from what it was before the pandemic~\cite{vyas2022new,howell2022coworking}.
Right now, for the foreseeable future, there is a large consensus among our informants that there will be a number of different work arrangements.
The first issue to mention is that organizations should account for individual differences between developers. 
Namely, when we performed cross-country comparisons, we had to acknowledge that the within-country variability outweighed the between-country variability.
This is a strong indicator that suggests high personality differences, that can not be captured with one-fits-all solutions.
In this regard, Smite et al., discuss the spectrum of different types of work arrangements based on two dimensions: work location (from fully remote to fully in-presence), and work schedule (i.e., fixed or flexible)~\cite{Smite2023workplace}.
However, to do so, a change in company culture and management support is an underlying assumption for our surveyed developers.
More in detail, work arrangements should depend not only on individual preferences, but also on the type of task that has to be performed (cf.~\cite{russo2021developers}).
For example, a development team that is relying on pair programming practices, might agree on a flexible location setting, but their work schedules have to be fully aligned. 
The last big challenge is the set up of face-to-face communication opportunities among colleagues.
Creating a sense of belonging and ownership is critical, especially for remote teams with a flexible schedule mode.
In this regard, embedding activities within the project lyfecicle aimed at ``re-socializing'' to align behaviors and norms needed to understand and work within a specific organization has proven effective~\cite{oshri2007global}.

\begin{table*}[!ht]
\centering
\small
\caption{Summary of key findings \& recommendations}
\label{tab:summary}
\begin{tabular}{@{}p{2cm}p{5.5cm}p{5.5cm}@{}}
\toprule
    & \textbf{Findings} & \textbf{Recommendations} \\
    \midrule
 
     Developers' well-being increased during the pandemic & Well-being consistently increased across all six time points, indicating that they bounced back from the negative impact the pandemic likely had on their well-being initially. & Developers showed a high level of resilience when working from home and improved their well-being. Software companies can extensively implement (hybrid) working from home practices. \\ \addlinespace


     Productivity remained unchanged & Developers' productivity has not changed across all six time points. & Working from home is not \textit{per se} detrimental for productivity. If organizations keep reasonable work expectations, professionals will be as productive at home as in the office. \\ \addlinespace


    Developers felt less lonely and improved their social contacts in general & This suggests that developers managed to reduce their loneliness, presumably by improving the quality and quantity of their social interactions.  & Active inclusion policies should be set in place for employees working from home. Mainly middle-management should focus on individual employees performance and their level of integration and communication with the team. \\ \addlinespace
    


    

      \\ \addlinespace


      No country difference (USA vs UK) when dealing with the pandemic & Our findings indicate that people living in the UK and the USA were impacted and `recovered' from the initial shock of the pandemic to a similar extent.  & Especially during another disastrous event, organizations can plan the same remote work strategies across both countries. \\ \addlinespace


      The Future of Work will be diverse. & When the company preferred employees to work from the office, developers who preferred working remotely were more likely to intent to change jobs. & Organizations need a flexible work arrangement solutions to retain and attract talents. \\ \addlinespace

    \bottomrule

\end{tabular}
\end{table*}

\subsection{Limitations}

In the following section, we discuss the most relevant limitations of this work.

\textit{Reliability}. For this investigation, we employed a six wave longitudinal design.
Informants have been identified through a multi-stage selection screening to ensure they were representative of the software engineering population.
Also, we computed an \textit{a priori} power analysis to identify the minimum number of participants required to provide reliable conclusions.
The internal consistencies (i.e., Cronbach's $ \alpha$) ranged from satisfactory to very good, with only very few exceptions such as disliking commuting.
 
\textit{Construct validity}. 
For this study we used $15$ variables previously identified in the literature that are related to well-being and productivity.
For any variable, we used a dedicated measurement instrument.
Construct validity was assessed by correlating all variables with each other, separately in each wave. The correlations were in the expected directions and in line with the literature~\cite{diener_beyond_2009,russo2020predictors,miller2011loneliness}. 
Our productivity measure was partly based on existing measures and showed some test-retest reliability. We are, however, limited in further validation of this self-perceived measure of productivity. 
We, therefore, encourage future work to consider additional validation of productivity, for example through other measures~\cite{del2011measuring} and analysis approaches (e.g., factor analysis).

\textit{Conclusion validity}.
We draw our conclusions based on a number of statistical analyses: Friedman's test, a non-parametric alternative to within-subject ANOVA, between-subject t-tests, linear mixed effects models, regressions, and correlations. We furthermore performed a thematic analysis.
To increase the trustworthiness of our results, we adjusted our alpha-thresholds to reduce the risk of false positives (i.e., Type I errors).
We acknowledge that some variations in our data might exist due to the fact that lockdown measures were not uniform in different countries.
To address this issue as best as possible, we only selected participants living in countries that during the first wave had similar regulations (we excluded, e.g., Sweden, Denmark).
We find very similar results when looking at between-country mean differences.
Moreover, some of our conclusions are based on general measures. For example, while we found that the quality of social contacts correlated positively with well-being and improved between waves 1 and 6, it is unclear which social contacts were mostly responsible for this. Thus, future research could directly ask for the quality of contact for various social relationships (e.g., with partner, family, friends, colleagues) and correlate the outcomes with well-being to get a better understanding of which social contact contribute most to developers' well-being.

\textit{Internal validity}.
Our study relies on self-reported measures, limiting the study's internal validity due to potential response biases.
Although our informants have been initially identified in other work~\cite{russo2020gender}, we applied several quality checks also after each time point.
Additionally, we searched for inaccurate or unlikely responses (of which we found none, which ensures data quality).
The attrition rate across the six waves is comparable to other longitudinal studies across a similar timespan~\cite{bardi2014value, feinberg2019understanding}.
Due to the evolving nature of the pandemic, data collection has been performed based on the information available at that point in time. 
As a consequence, the time spans are not homogeneous but represent moments of the pandemic where data collection seemed to be representative of the pandemic trend, potentially affecting data variability.
Finally, we note that by aiming to limit the survey length to reduce participant attrition may have resulted in relevant variables that could have explained some of our results not being included.
Future work may include additional variables as indicated to be relevant by prior work or theory. 

\textit{External validity}.
The primary aim of our longitudinal analysis was to maximize internal validity by finding significant effects.
Our recruitment through Prolific cannot be guaranteed to be representative of the whole software engineering population.
Further, we limited our sample to software developers who worked for at least 20 hours a week, which limits the generalisability of our findings to software developers who are at least working part-time. There might have been an unknown number of developers whose work-related productivity substantially declined because of health issues or caring responsibilities, for example. 
Additionally, this study faces certain limitations due to the heterogeneity of pandemic policies across countries considered in our analysis. Our analysis was based on the initial information available in April 2020, which may not accurately reflect the subsequent policy changes and adaptations that occurred as the situation evolved.

\section{The Integrated Job Demands-Resources and Self-Determination Model (IJARS)}
\label{sec:theory}

Building upon our findings and the existing literature, we propose the Integrated Job Demands-Resources and Self-Determination Model (IJARS) as a comprehensive framework to explain the well-being and productivity of software engineers during the COVID-19 pandemic, as summarized in Figure~\ref{fig:IJARS}. The IJARS model combines elements from both the Job Demands-Resources (JD-R) Model and Self-Determination Theory (SDT) to examine how job demands and resources interact with the satisfaction of basic psychological needs (autonomy, competence, and relatedness) in the context of remote work.

\subsection{Job Demands, Resources, and Basic Psychological Needs}

The Job Demands-Resources Model posits that job demands and resources have a combined impact on employees' well-being and productivity~\cite{bakker2007job}. Job demands refer to aspects of a job that require effort, while job resources are those aspects that help individuals achieve work goals, reduce job demands, or stimulate personal growth~\cite{demerouti2001job}. In the IJARS model, we propose that job demands and resources can potentially influence the satisfaction or frustration of the basic psychological needs posited by Self-Determination Theory~\cite{ryan2000self}. For instance, high job demands, such as workload or role ambiguity, could undermine the satisfaction of competence or autonomy, leading to decreased well-being and productivity.

Conversely, job resources, such as management support or opportunities for skill development, could enhance the satisfaction of basic psychological needs, thereby promoting well-being and productivity. This intersection of the JD-R model and SDT provides a more nuanced understanding of the factors that contribute to software engineers' well-being and productivity during the COVID-19 pandemic.
To better understand it, we are now briefly illustrating these two theories.

\subsubsection{The Job Demands-Resources Model}

The Job Demands-Resources (JD-R) model is an influential and widely applied theoretical framework within the field of occupational health psychology and organizational behavior~\cite{bakker2023job}. Developed by Demerouti et al.~\cite{demerouti2001job}, the JD-R model seeks to understand the complex interplay between job demands, job resources, and their impact on employee well-being and performance. The model posits that work environments can be characterized by job demands, which require sustained physical or mental effort, and job resources, which facilitate the achievement of work goals, reduce job demands, and stimulate personal growth and development~\cite{bakker2008job, bakker2017job}.
The key components of the JD-R Model are the following:

\begin{enumerate}
\item \textbf{Job Demands:} Job demands encompass the physical, mental, interpersonal, or institutional elements of a job that necessitate continuous exertion or expertise and can result in specific physiological or psychological consequences~\cite{demerouti2001job}. Instances of workplace requirements consist of task volume, deadline stress, role uncertainty, and emotional strain~\cite{bakker2008job}.

\item \textbf{Job Resources:} Job resources are the physical, psychological, social, or organizational aspects of a job that help employees achieve work goals, reduce job demands, or stimulate personal growth and development~\cite{demerouti2001job}. Examples of job resources include autonomy, social support, performance feedback, and opportunities for professional development~\cite{bakker2008job}.
\end{enumerate}

The JD-R model proposes two parallel processes that explain the relationship between job demands, job resources, and employees' well-being and performance~\cite{bakker2008job}:

\textit{Health Impairment Process:} This process posits that high job demands can lead to chronic work stress, which may result in exhaustion, burnout, and other negative health outcomes~\cite{demerouti2001job, bakker2008job}.

\textit{Motivational Process:} In contrast, the motivational process suggests that job resources can enhance work engagement, job satisfaction, and positive work outcomes such as increased performance and reduced turnover intentions~\cite{bakker2008job, bakker2017job}.

The JD-R model offers valuable insights for organizations and managers aiming to improve employee well-being and performance. By identifying and addressing critical job demands and enhancing job resources, organizations can mitigate the negative consequences of work stress and promote employee engagement, motivation, and success~\cite{bakker2008job}. Interventions based on the JD-R model may include job redesign, workload management, and the provision of social and instrumental support to employees~\cite{bakker2017job}.

\subsubsection{Self-Determination Theory}

Self-Determination Theory (SDT) is a macro theory of human motivation that has garnered significant attention and empirical support within the field of organizational behavior and management. Developed by Deci \& Ryan~\cite{deci1985intrinsic, deci2000what}, SDT is grounded in the belief that humans have an inherent propensity for growth, psychological well-being, and autonomous self-regulation. The theory posits that the satisfaction of three basic psychological needs -- autonomy, competence, and relatedness -- is essential for high motivation, performance, and well-being in various domains, including the workplace~\cite{ryan2000self, gagne2005self}.
The three basic psychological needs are highlighted as follows:

\begin{enumerate}
\item \textbf{Autonomy:} Autonomy refers to the need for individuals to feel a sense of volition and choice in their actions~\cite{deci2000what}. When people perceive their behaviors as self-directed and in accordance with their values and interests, they are more likely to experience intrinsic motivation, positive affect, and enhanced well-being~\cite{ryan2006self}.

\item \textbf{Competence:} Competence is the need to feel effective and capable in achieving desired outcomes~\cite{white1959motivation}. SDT posits that individuals are more intrinsically motivated when they feel competent and are given opportunities to develop and demonstrate their abilities~\cite{deci2000what, ryan2006self}.

\item \textbf{Relatedness:} Relatedness refers to the need to feel connected and cared for by others~\cite{baumeister1995need}. According to SDT, individuals are more likely to thrive when they experience a sense of belonging and support within their social environment~\cite{deci2000what, ryan2006self}.
\end{enumerate}

SDT offers valuable insights for managers seeking to foster employee motivation, engagement, and well-being. Research has demonstrated that employees who experience greater need satisfaction in their jobs exhibit higher levels of intrinsic motivation, job satisfaction, and performance~\cite{gagne2005self, vandenbroeck2016review}. As such, organizations that support employee autonomy, competence, and relatedness are likely to yield positive outcomes, such as increased productivity, reduced turnover, and enhanced overall performance~\cite{ryan2000self}.

\subsection{The Rational of the IJARS Model}

The IJARS model highlights the importance of adaptation and resilience in understanding changes in well-being and productivity over time, beyond the key variables of the JD-R model and SDT. Adaptation refers to the process of adjusting to new or altered circumstances, while resilience refers to the capacity to bounce back from adversity or maintain well-being and productivity in the face of challenges~\cite{hartmann2020resilience, luthar2000construct}. During the COVID-19 pandemic, software engineers faced numerous changes in their work environment, including shifting job demands and resources associated with remote work. The IJARS model posits that these professionals may have adapted to these changes and developed resilience~\cite{finstad2021resilience}, leading to improvements or stabilization in their well-being and productivity.

The COVID-19 pandemic has led to a significant shift in work environments across various industries, with remote work becoming the new normal for many professionals. For software engineers, this shift has resulted in changes in job demands and resources, including increased workload, limited social interaction, and reduced access to physical resources, such as office equipment and face-to-face support from colleagues and supervisors. These changes have created new challenges for software engineers, impacting their well-being and productivity.

To address these challenges, the IJARS model proposes that software engineers may engage in adaptive processes and develop resilience, which can help them maintain or even improve their well-being and productivity over time. Adaptive processes involve seeking out new resources, developing new skills, and finding new ways to satisfy basic psychological needs, such as autonomy, competence, and relatedness. These processes can help software engineers cope with the changes in their work environment and manage the challenges associated with remote work.

Resilience, on the other hand, refers to the capacity to bounce back from adversity or maintain well-being and productivity in the face of challenges~\cite{hartmann2020resilience, luthar2000construct}. Resilience is not a static trait but rather a dynamic process that can evolve over time~\cite{finstad2021resilience}. In the context of the COVID-19 pandemic, resilient software engineers may have adapted to the new job demands and resources, leveraged available resources, and satisfied their basic psychological needs despite the disruptions caused by the pandemic. By doing so, they may have been better able to maintain their well-being and productivity, even under difficult circumstances.

The rationale of the IJARS model is based on the premise that software engineers' well-being and productivity are not solely determined by job demands and resources but also by their ability to adapt and develop resilience in response to changes in their work environment. The model proposes that by understanding the interaction between job demands, resources, and basic psychological needs, as well as the role of adaptation and resilience, organizations can better support software engineers' well-being and productivity during the COVID-19 pandemic and beyond.

\subsubsection{The Role of Resilience in the IJARS Model}

Resilience plays a crucial role in the IJARS model in understanding the well-being and productivity of software engineers during the COVID-19 pandemic. Resilience refers to the capacity to recover from adverse events or maintain well-being and productivity in the face of challenges~\cite{masten2001ordinary}. In the context of this study, software engineers faced numerous changes in their work environment due to the pandemic, such as the shift to remote work, increased reliance on virtual communication, and the need to balance work and personal life under the same roof.

In our longitudinal analysis, we found that software engineers demonstrated a high level of resilience in coping with the challenges posed by the pandemic. Despite the initial decline in well-being and productivity reported by some studies~\cite{Ralph2020pandemic}, our results indicate that well-being increased and productivity remained stable over time. This resilience can be attributed to several factors, including the ability to adapt to remote work, the maintenance of social contacts, and the development of new coping strategies in response to stressors~\cite{tugade2004resilient,luthar2000construct}. The resilience of software engineers is further evidenced by their ability to bounce back and maintain their well-being and productivity even when lockdown measures were partially lifted.

Our findings support the notion that resilience is a key factor in understanding the well-being and productivity of software engineers during challenging times. Although we have not found an association between resilience and productivity, other research has with other productivity measures~\cite{shatte2017positive, zehir2016effects}, suggesting that the way how productivity is measured impacts the outcome. It is crucial for organizations to recognize the importance of resilience in their workforce and to foster a supportive environment that encourages its development~\cite{avey2009psychological}.

\subsubsection{The Role of Adaptation in the IJARS Model}

Adaptation, defined as the process of adjusting to new or altered circumstances~\cite{lazarus1984stress}, is another vital component of the IJARS model. In the context of the COVID-19 pandemic, software engineers had to adapt to various changes in their work environment, including remote work, changes in job demands and resources, and navigating new communication technologies.

Our longitudinal analysis showed that software engineers were able to adapt to these changes, as evidenced by the increase in well-being and the stable levels of productivity observed over the 24-month study period. The ability to adapt was facilitated by several factors, such as the implementation of new work practices, the improvement of communication channels, and the development of strategies to manage work-life balance in a remote work setting~\cite{storey2017social,meyer2019developers}.

Moreover, the adaptation process was not uniform across individuals, with some software engineers finding remote work more challenging than others. This highlights the importance of understanding individual differences in the adaptation process and tailoring support accordingly~\cite{gajendran2007good}.

In summary, adaptation plays a pivotal role in the IJARS model, helping to explain the well-being and productivity of software engineers during the COVID-19 pandemic. Organizations should consider the importance of adaptation and promote a supportive environment that enables employees to adjust to new circumstances and thrive in the face of change~\cite{conway2002daily}.

\subsection{Integrating SDT, Job Demand, Job Resources, Adaptation, and Resilience for Well-Being and Productivity}

The Integrated Job Demand-Resources Self-Determination (IJARS) model brings together the concepts of Self-Determination Theory (SDT), the Job Demands-Resources (JD-R) model with Adaptation and Resilience to provide a comprehensive understanding of their effects on well-being and productivity in the workplace.

\begin{figure}[h]
\centering
\includegraphics[width=1\linewidth]{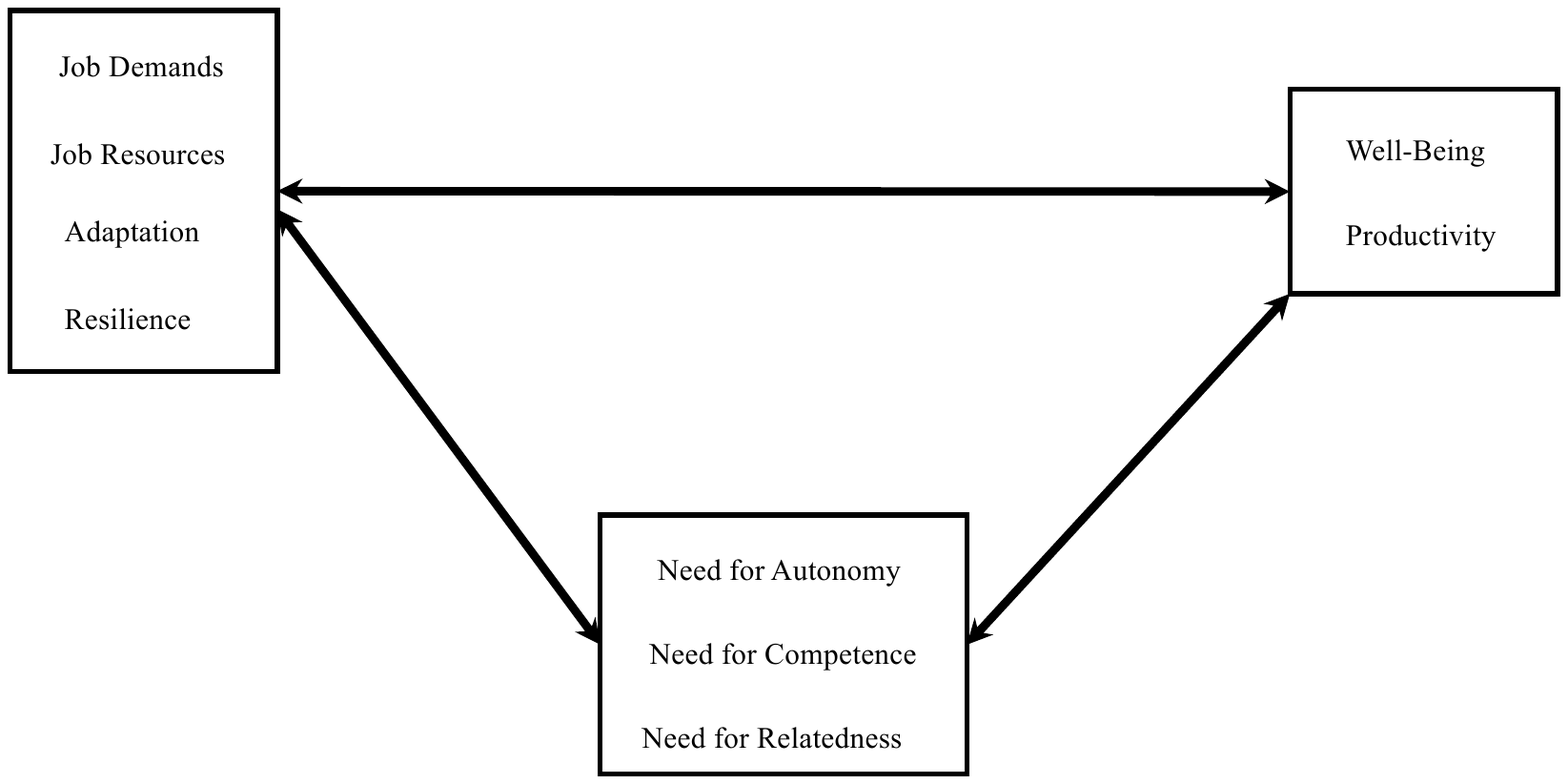}
\caption{The Integrated Job Demands-Resources and Self-Determination Model.}
\label{fig:IJARS}
\end{figure}

The IJARS model underscores the intricate interplay between job demands and job resources, and how these interactions can shape the satisfaction of three fundamental psychological needs — autonomy, competence, and relatedness — which lie at the core of Self-Determination Theory (SDT)~\cite{deci2011self}. These needs serve as critical mediators, linking the dynamics between job demands, job resources, adaptation, resilience, and the ultimate outcomes, namely well-being and productivity. An illustrative depiction of the model can be seen in Figure~\ref{fig:IJARS}.

Delving deeper into the dynamics, job demands, which might encompass high work pressure, tight deadlines, or complex tasks, could potentially exert a negative influence on well-being and productivity. This is primarily because they can compromise the satisfaction of basic psychological needs, leading to stress and burnout~\cite{bakker2007job}.

However, job resources can effectively counterbalance the adverse effects of these job demands. These resources could manifest as tangible elements, such as appropriate equipment or a conducive work environment, or intangible aspects like social support, feedback, or opportunities for professional development. These resources can bolster the satisfaction of autonomy, competence, and relatedness~\cite{bakker2008job}, which are essential to thriving in the workplace. When employees are provided with an environment that fosters these needs, they are likely more resilient in the face of high job demands. For instance, with adequate social support or feedback, employees can navigate challenging job demands more effectively, thereby experiencing higher levels of well-being and productivity. Importantly, we acknowledge that the relations between the three needs as well as job demands, job resources, adaptation, and resilience are likely \textbf{bidirectional}~\cite{xanthopoulou2009reciprocal}. For example, while job demands can impact autonomy~\cite{ng2008personality}, software developers may choose their job based on their need for autonomy. Similarly, resilience may increase people's needs for autonomy, competence, and relatedness~\cite{trigueros2019influence}, the opposite pattern is also likely true~\cite{liu2021effects}.  

Adaptation and resilience also play a vital role in the IJARS model. The ability to adapt to changing job demands and resources is essential for maintaining or enhancing the satisfaction of basic psychological needs~\cite{lazarus1984stress}. As a result, individuals who successfully adapt are more likely to experience better well-being and productivity~\cite{meyer2014software}. Likewise, resilience enables employees to effectively cope with and recover from adversity, which contributes to the maintenance or regaining of need satisfaction and, consequently, promotes well-being and productivity~\cite{masten2001ordinary, tugade2004resilient}.

It is important to note that the associations between all variables are likely going to be bi-directional. For example, while research, including ours, tends to use well-being and productivity as outcome variables and argues that both variables are influences by resilience, job demands, and so on ~\cite{Ralph2020pandemic,russo2020predictors,van2010time}, we argue that it is conceivable that both variables also impact the other variables of the IJARS model. For instance, productivity might impact resilience through increased self-efficacy: If a person feel productive, it helps them to believe more in themselves which in turn increases their resilience in case they have an unproductive week (e.g., ``unproductive week? Doesn't matter. I was productive in the past and can be productive next week again''). Indeed, there is sparse evidence from longitudinal studies that productivity can be a predictor of variables such as goal or service orientation, and general health~\cite{geibel2022all,van2014relationships}. 

SDT serves as the central mediator in the IJARS model by demonstrating how the satisfaction of basic psychological needs links job demands, job resources, adaptation, and resilience to well-being and productivity. By attending to the satisfaction of these needs, organizations can create a supportive environment that fosters adaptive coping strategies and resilience, ultimately promoting well-being and productivity in the face of diverse job demands and resources.

In conclusion, the IJARS model offers a comprehensive framework for understanding the complex relationships between job demands, job resources, adaptation, resilience, and the satisfaction of basic psychological needs within SDT, and how these factors contribute to well-being and productivity in the workplace. By focusing on the interactions between these concepts and the central role of SDT, the IJARS model provides valuable insights for both researchers and practitioners in the fields of organizational behavior, human resources, and management.

\subsection{Implications and Applications}

The IJARS model offers several implications for both theory and practice. By integrating SDT with job demand, job resources, adaptation, and resilience, the model contributes to a more comprehensive understanding of the factors that influence well-being and productivity in the context of a lockdown. 

\subsubsection{Organizational Strategies}

Organizations can use the IJARS model to identify areas where interventions may be beneficial. For example, they can assess job demands and resources to determine if any imbalances exist and address them accordingly. Implementing strategies such as job redesign or providing additional resources can help alleviate excessive job demands, ultimately leading to improved well-being and productivity~\cite{bakker2014job}.
Moreover, fostering a supportive organizational culture that encourages collaboration and open communication can enhance employees' sense of belonging, leading to increased adaptation and resilience in the face of stressors~\cite{storey2017social}. Additionally, promoting individual and team autonomy, competence, and relatedness may further support self-determination, enhancing employees' intrinsic motivation and overall well-being~\cite{gagne2005self}.

\subsubsection{Employee Development and Training}

The IJARS model can also guide employee development and training programs. Focusing on building resilience skills, such as problem-solving, emotional regulation, and adaptability, can better equip employees to manage workplace challenges~\cite{masten2001ordinary}. Furthermore, training programs that emphasize self-awareness and self-regulation can foster employees' abilities to recognize and manage their needs, facilitating adaptive responses to job demands and resources~\cite{tugade2004resilient}.

\subsubsection{Performance Management}

Performance management systems can benefit from the IJARS model by considering the interconnectedness of job demands, resources, adaptation, resilience, and self-determination in evaluating employee performance. By incorporating these factors into performance appraisals and feedback, organizations can better understand the underlying influences on employee performance and provide more targeted support and development opportunities~\cite{meyer2019developers}.

In conclusion, the IJARS model offers a holistic perspective on the interplay between job demands, resources, adaptation, resilience, self-determination, well-being, and productivity. By incorporating this model into organizational strategies, employee development, and performance management, organizations can better support their employees and create a thriving working from home environment during a lockdown.

\section{Conclusion}
\label{sec:conclusion}

Over a two-year period from April 2020 to April 2022, our groundbreaking six-wave longitudinal study delved into the lives of 192 software developers amidst the COVID-19 pandemic, examining the evolution of their well-being, productivity, and 15 associated social and psychological factors. We also contrasted the experiences of developers based in the USA and the UK.

Our analysis revealed increases in well-being, and quality of social contacts, coupled with a decrease in emotional loneliness. Moreover, the location of the developers (USA or UK) did not yield any significant differences across all six data collection waves. We discovered in wave 6 that preference for remote work bore no significant correlation with productivity, well-being, solidarity towards the company, or job satisfaction, signifying that developers who prefer working from home are equally productive, solidary, and satisfied with their job compared to their office-based counterparts. Interestingly, while job satisfaction and company solidarity were both negatively correlated with the intention to change jobs, boredom, and loneliness, they were positively associated with well-being and needs, but showed no relationship with productivity.

Our study's robustness lies in its extensive time frame, meticulous informant selection following an \textit{a priori} power analysis, and careful alpha level adjustment to mitigate false-positive results and misleading recommendations. This research represents the most comprehensive longitudinal analysis of software engineers amidst the COVID-19 pandemic, offering valuable insights into the software engineering community's WFH policies and illuminating the future of work post-pandemic.

We introduce the Integrated Job Demands-Resources and Self-Determination Model (IJARS) as a comprehensive framework for understanding the well-being and productivity of software engineers during the pandemic, melding the Job Demands-Resources (JD-R) Model and Self-Determination Theory (SDT) to assess the interplay between job demands, resources, and the fulfillment of basic psychological needs in remote work contexts.

\textbf{Future research} should prioritize long-term assessments of software engineers' work conditions and arrangements, as well as the broader implications of our findings. In particular, more in-depth investigations are needed to explore the impact of hybrid and flexible working arrangements on software professionals' well-being, productivity, and the subsequent effects on software projects. Such research could include examining the influence of remote work on team dynamics, communication, skill development, and organizational culture.

Specifically, an interesting opportunity for future research lies in the study of fluid teams, which involve flexible team compositions that evolve over time in response to project needs. Investigating how fluid teams function and adapt in remote work environments could provide valuable insights into the challenges and benefits of such structures. It is essential to understand how fluid teams affect software engineers' well-being, productivity, and job satisfaction, as well as the overall success of software projects.

Additionally, future studies should aim to validate and refine the IJARS model by examining its applicability in diverse settings and industries beyond software engineering. This would help establish the model's generalizability and provide further insights into the interplay between job demands, resources, and basic psychological needs in different work contexts. Moreover, research could assess the effectiveness of interventions or strategies based on the IJARS model in promoting employee well-being, productivity, and overall job satisfaction.

Lastly, it would be valuable to investigate the long-term mental health effects of remote work on software engineers and identify potential support mechanisms for employees working in isolation. Understanding these effects could contribute to the development of targeted interventions that maintain mental well-being and foster a healthy work environment in remote and hybrid settings. By pursuing these research opportunities, we can expand our understanding of remote work and its long-term effects on the software engineering profession, helping organizations adapt to the evolving landscape and foster the well-being and productivity of their employees.


\section*{Supplementary Materials} 
The complete replication package is openly available under a CC BY 4.0 license on Zenodo, DOI: \url{https://doi.org/10.5281/zenodo.7828605}.


\section*{Acknowledgment}
This work was supported by the Carlsberg Foundation under grant agreement number CF20-0322 (PanTra --- Pandemic Transformation).

\bibliographystyle{ACM-Reference-Format}
\bibliography{bib}

\newpage
\begin{appendix}

\section{Competency-based questions}
\label{sec:appendixCompetence}

Appendix \ref{sec:appendixCompetence} reports all the questions used to assess participants' developer competencies.

\vspace{2em}

\noindent
\underline{\hspace{2.5cm}} is a measure of the degree of interdependence between modules.

\begin{itemize}
    \item Cohesion
    \item Coupling
    \item None of the mentioned
    \item All of the mentioned
\end{itemize}
 
\noindent
What are the advantages of arrays?

\begin{itemize}
    \item Objects of mixed data types can be stored
    \item Elements in an array cannot be sorted
    \item Index of first element of an array is 1
    \item Easier to store elements of same data type
\end{itemize}
 
\noindent
Which of the following best defines a class?
\begin{itemize}
    \item Parent of an object
    \item Implementation of an object
    \item Blueprint of an object
    \item Scope of an object
\end{itemize}

\newpage

\section{Questionnaire instruments}
\label{sec:appendixQuestions}

Appendix \ref{sec:appendixQuestions} reports all the questionnaires used in all six waves.

\subsection{Well-being -- Satisfaction with Life Scale~\cite{diener1985satisfaction} -- Waves 1-6}
\noindent
Below are five statements that you may agree or disagree with. Please be open and honest in your responding. [7-point response scale ranging from Strongly disagree -- Strongly agree]

\begin{itemize}
\item I was satisfied with my life in the past week.
\item The conditions of my life in the past week were excellent.
\item If I could live the past week over again, I would change almost nothing.
\item In the past week, I have gotten the important things I want.
\item In most ways, my life in the past week has been close to my ideal.
\end{itemize}

\subsection{Productivity -- Self-developed scale -- Waves 1-6}

\noindent
Please answer the following questions about your work. Remember that all answers are anonymous.

\begin{itemize}
\item How many hours have you been working approximately in the past week? [0--80 hours]
\item How many hours were you expecting to work over the past week assuming there would be no global pandemic and lockdown? [0--80 hours]
\end{itemize}

\begin{itemize}
\item How many tasks that you were supposed to complete last week did you effectively manage to complete? [0--100\%]
\end{itemize}

\begin{itemize}
\item If you rate your productivity (i.e., outcome) per hour, has it been more or less over the past week as compared to a normal week? [100\% less productive -- 100\% or more productive]
\end{itemize}

\subsection{Boredom -- Boredom Proneness Scale~\cite{farmer1986boredom,struk2017short} -- Waves 1-6}

\noindent
Please indicate to what extent you agree with the following statements. [7-point response scale ranging from Strongly disagree -- Strongly agree]

\begin{itemize}
\item In most situations, it is hard for me to find something to do or see to keep me interested.
\item Much of the time, I just sit around doing nothing.
\item It takes more stimulation to get me going than most people.
\item I often find myself at “loose ends,” not knowing what to do.
\item I don’t feel motivated by most things that I do.
\item Many things I have to do are repetitive and monotonous.
\item Unless I am doing something exciting, even dangerous, I feel half-dead and dull.
\item I find it hard to entertain myself.
\end{itemize}

%
%
%
%
%
%
\subsection{Self-blame and behavioral disengagement -- Subscales of the Brief COPE scale~\cite{Carver1997BriefCOPE} -- Waves 1-6}

\noindent
These items deal with ways you've been coping with the stress in your life \underline{in the past week}. There are many ways to try to deal with problems. Obviously, different people deal with things in different ways, but we are interested in how you've tried to deal with it. Use these response choices. Try to rate each item separately in your mind from the others. Make your answers as true FOR YOU as you can. [5-point response scale ranging from I've not been doing this at all -- I've been doing this a lot]

\begin{itemize}
\item I've been giving up trying to deal with it. [Behavioral disengagement]
\item I’ve been criticizing myself. [Self-blame]
\item I've been giving up the attempt to cope. [Behavioral disengagement]
\item I’ve been blaming myself for things that happened. [Self-blame]
\end{itemize}

\subsection{Autonomy, competence, and relatedness -- Psychological needs scale~\cite{sheldon2012balanced} -- Waves 1-6}

\noindent
Please read each of the following statements carefully, thinking about how true it was for you \underline{in the past week}. [5-point response scale ranging from No agreement -- Much agreement]

\begin{itemize}
\item I took on and mastered hard challenges.
\item I experienced some kind of failure, or was unable to do well at something.
\item There were people telling me what I had to do.
\item I had a lot of pressures I could do without.
\item I felt a strong sense of intimacy with the people I spent time with.
\item I felt unappreciated by one or more important people.
\item I was free to do things my own way.
\item I was successfully completing difficult tasks and projects.
\item I had to do things against my will.
\item I did something stupid, that made me feel incompetent.
\item I had disagreements or conflicts with people I usually get along with.
\item I felt close and connected with other people who are important to me.
\item I felt a sense of contact with people who care for me, and whom I care for.
\item I was really doing what interests me.
\item I was lonely.
\item I struggled doing something I should be good at.
\item I did well even at the hard things.
\end{itemize}

%
%
%

\subsection{Quality and quantity of communication with colleagues and line managers -- Self-developed scale -- Waves 1-6}

\noindent
The following questions refer to communication with colleagues and line managers. If you don’t have any colleagues or line managers, please skip the following three items. [6-point response scale ranging from Strongly disagree -- Strongly agree]

\begin{itemize}
\item I feel that my colleagues and line manager believed in me over the past week.
\item I feel that my colleagues and line manager have been supporting me over the past week.
\item Overall, I am happy with the interactions with my colleagues and line managers over the past week.
\end{itemize}




\subsection{Stress -- Perceived Stress Scale~\cite{cohen1983global} -- Waves 1-6}

\noindent
The questions in this scale ask you about your feelings and thoughts during the \underline{last week}. [5-point response scale ranging from Never -- Very often]

\begin{itemize}
\item In the last week, how often have you felt that you were unable to control the important things in your life?
\item In the last week, how often have you felt confident about your ability to handle your personal problems?
\item In the last week, how often have you felt difficulties were piling up so high that you could not overcome them?
\item In the last week, how often have you felt that things were going your way?
\end{itemize}

%
%
%
%
%
%
%
%

\subsection{Extraversion -- Subscale of the Brief HEXACO Inventory~\cite{DeVries2013HEXACO} -- Waves 1-6}

\noindent
Please indicate to what extent you agree with the following statements. [5-point response scale ranging from Strongly disagree -- Strongly agree]

\begin{itemize}
\item I like to talk with others.
\item I easily approach strangers.
\item Nobody likes talking with me.
\item I am seldom cheerful.
\end{itemize}

%
%
%
\subsection{Distractions at home -- Self-developed scale -- Waves 1-6}

\noindent
Distractions at home [5-point response scale ranging from Not at all -- Very often]

\begin{itemize}
\item I am often distracted from my work (e.g., noisy neighbors, children who need my attention)
\item I am able to focus on my work for longer time periods
\end{itemize}

%
%
%

\subsection{Generalized anxiety -- adapted version of the $7$-item Generalized Anxiety Disorder scale~\cite{Spitzer2006GAD7} -- Waves 1-6}

\noindent
Over the \underline{last week}, how often have you been bothered by the following problems? [5-point response scale ranging from Not at all -- Every day]

\begin{itemize}
\item Feeling nervous, anxious or on edge.
\item Not being able to stop or control worrying.
\item Worrying too much about different things.
\item Trouble relaxing.
\item Being so restless that it is hard to sit still.
\item Becoming easily annoyed or irritable.
\item Feeling afraid as if something awful might happen.
\end{itemize}

\subsection{Pandemic concerns -- Self-developed scale -- Wave 1-6}
\noindent
Over the \underline{last week}, have you been concerned about the following problem? [5 steps, Not at all concerned -- Extremely concerned]

\begin{itemize}
\item How concerned do you feel about COVID-19?
\item How concerned do you feel about future pandemics?
\end{itemize}

\subsection{Emotional and social loneliness -- De Jong Gierveld Loneliness Scale~\cite{gierveld2006} -- Waves 1-6}

\noindent
Over the \underline{last week}, how much do the following statements apply to you? [5-point response scale ranging from Not at all -- Every day]

\begin{itemize}
\item I experience a general sense of emptiness.
\item I often feel rejected.
\item I miss having people around.
\item There are plenty of people I can rely on when I have problems.
\item There are enough people I feel close to.
\item There are many people I can trust completely.
\end{itemize}

%
%
%
%

\subsection{Quality of social contacts -- Two items adapted from the social relationship quality scale~\cite{birditt2007relationship} and one self-developed item -- Waves 1-6}
\noindent
The following questions refer to your social contacts outside of work. [5-point response scale ranging from Strongly disagree -- Strongly agree]

\begin{itemize}
\item I feel that the people with whom I have been in contact over the past week support me.
\item I feel that the people with whom I have been in contact over the past week believe in me.
\item I am happy with the amount of social contact I had in the past week. 
\end{itemize}

\subsection{Demographics}
\noindent
Demographics and debriefing
\noindent
You almost made it! Now some questions about yourself.

\noindent
What is your gender? [single-selection]

\begin{itemize}
\item Woman
\item Man
\item Non-binary
\item Prefer not to disclose
\item Prefer to self-describe [...]
\end{itemize}

\noindent
In which country are you currently based? [single-selection]

\begin{itemize}
\item United Kingdom
\item United States
\item Other [...]
\end{itemize}

\noindent
In which state do you currently reside? [dropdown]

\noindent
Is there still a lockdown where you are living (i.e., are still all schools and non- essential shops closed)? [single-selection]

\begin{itemize}
\item Yes
\item Unsure
\item No
\end{itemize}

\noindent
How old are you? [in years, input field]

\noindent
My living situation at the moment: [single-selection]

\begin{itemize}
\item Living alone
\item Living with other people
\end{itemize}

\noindent
How many of the people you're living with at the moment are: [input boxes]

\begin{itemize}
\item Babies / Infants (0--1 years old)
\item Toddlers (1--3 years old)
\item Children (4--11 years old)
\item Teenagers (12--17 years old)
\item Adults (18+ years old)
\end{itemize}

\noindent
What type of organization do you work for?

\begin{itemize}
\item Public
\item Private
\item Other
\item Unsure
\end{itemize}

\noindent
What was your approximate yearly household income before taxes in US-Dollar in 2019?

\begin{itemize}
\item <20,000
\item 20,000-40,000
\item 40,001-60,000
\item 60,001-80,000
\item 80,001-100,000
\item >100,000
\end{itemize}

\noindent
What percentage of your time have you been working remotely (i.e., not physically in your office) over the past 12 months? [input field]

%

\noindent
Thank you for participating in the first wave of this longitudinal study. We will contact you again in approximately one week and in two weeks and ask you to complete a shorter survey. It is important for us that you participate in all three waves.

\noindent
Do you have any comments so far? [freetext input]

\section{Questionnaire instruments -- Wave 6}
\label{sec:appendixQuestionsW6}
Appendix \ref{sec:appendixQuestionsW6} reports the additional questionnaire used exclusively in wave 6.

\subsection{Preference for working from home -- Self-developed scale -- Wave 6}
Additional questions

\begin{itemize}
    \item  Do you prefer working from home or from the office in the subsequent years? [0--100 ranging from Want to go to the office to Want to continue working from home]
    \item If I could choose myself, I would work [0--100 ranging from In my office full-time to From anywhere full-time]
    \item My company allows me to: [0--100 ranging from Go back to the office fulltime to Working from anywhere full-time, with option `No clear policy from my company yet on this regard']
    \item I expect to: [0--100 ranging from Go back to the office fulltime to Working from anywhere full-time, with option `No clear policy from my company yet on this regard']
\end{itemize}

\subsection{Resilience -- Brief Resilience scale~\cite{smith2008brief} -- Wave 6}
Please indicate the extent to which you agree with each of the following statements [5-point response scale ranging from Strongly disagree to Strongly agree]

\begin{itemize}
    \item I tend to bounce back quickly after hard times
    \item I usually come through difficult times with little trouble
    \item It is hard for me to snap back when something bad happens
    \item I tend to take a long time to get over set-backs in my life
    \item It does not take me long to recover from a stressful event
    \item I have a hard time making it through stressful events
    \item I confirm that I pay attention and select disagree
\end{itemize}

\subsection{Solidarity with the company -- 3-item Solidarity subscale of the In-group Identification Scale~\cite{leach2008group} -- Wave 6}
Below are a few questions about how you feel towards your company [5-point response scale ranging from Strongly disagree to Strongly agree]

\begin{itemize}
    \item I feel a bond with my company
    \item I feel solidarity with my company
    \item I feel committed to my company
\end{itemize}

\subsection{Job satisfaction-- 4 highest loading items of the Generic Job Satisfaction Scale~\cite{macdonald1997generic} -- Wave 6}
Below are a few questions about how you feel towards your company [5-point response scale ranging from Strongly disagree to Strongly agree]

\begin{itemize}
    \item I feel good about my job
    \item I receive recognition for a job well done
    \item l feel secure about my job
    \item I feel good about working at this company
\end{itemize}

\subsection{Commuting experience -- Self-developed scale -- Wave 6}

\begin{itemize}
    \item I find commuting to work effortful
    \item What I like about working from home is that I do not have to commute
\end{itemize}

\subsection{Intention to change jobs -- Self-developed scale -- Wave 6}
Please respond to the following statements [7-point response scale ranging from Not at all -- Very]

\begin{itemize}
    \item I am likely to change jobs within the next year
    \item I am considering changing jobs soon
\end{itemize}

\subsection{Changed jobs -- Self-developed scale -- Wave 6}
Have you changed jobs since March 2020?

\begin{itemize}
    \item No
    \item Yes, once
    \item Yes, twice
    \item Yes, more than twice
    \item I became unemployed
\end{itemize}

\noindent
My decision to change jobs was affected by my company's work from home policy.

\begin{itemize}
    \item No, this did not play a role.
    \item Yes, I wanted to work more from home than my company allowed.
    \item Yes, I wanted to work more at the office than my company allowed
\end{itemize}

\section{Changes along COVID-19 Pandemic}
\label{sec:appendixC}

Appendix \ref{sec:appendixC} reports additional figures for our Friedman's test (changes across the two years).

\begin{figure}[h]
\centering
\includegraphics[width=0.8\linewidth]{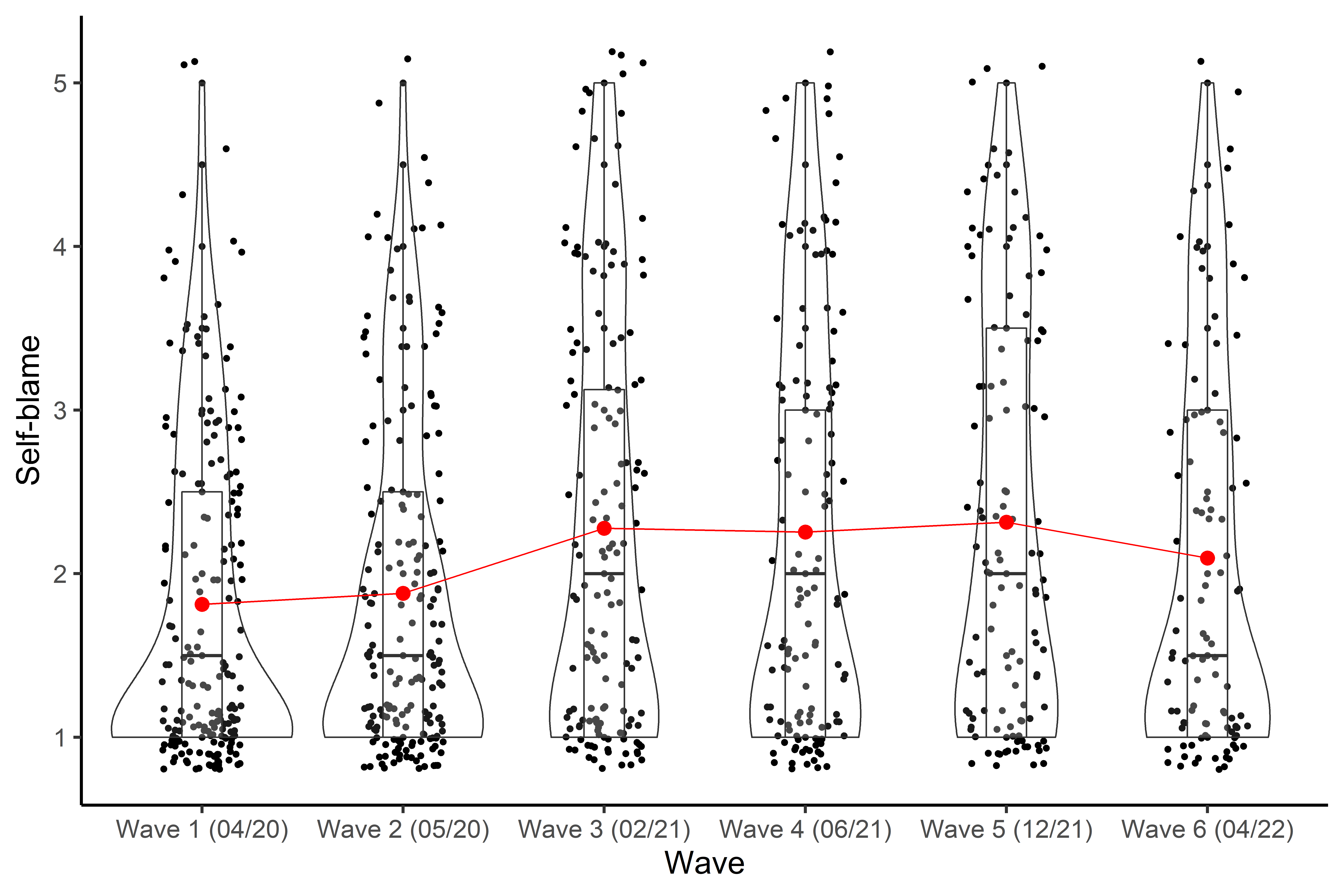} 
\caption{Self-blame across time. The red line displays the trend over time, whereas the box at each time point shows the range in which the middle 50\% of the data falls. Responses were given on a 5-point scale, with 1 being the lowest possible score and 5 being the highest possible score.}
\label{fig:sb}
\end{figure}

\begin{figure}[h]
\centering
\includegraphics[width=0.8\linewidth]{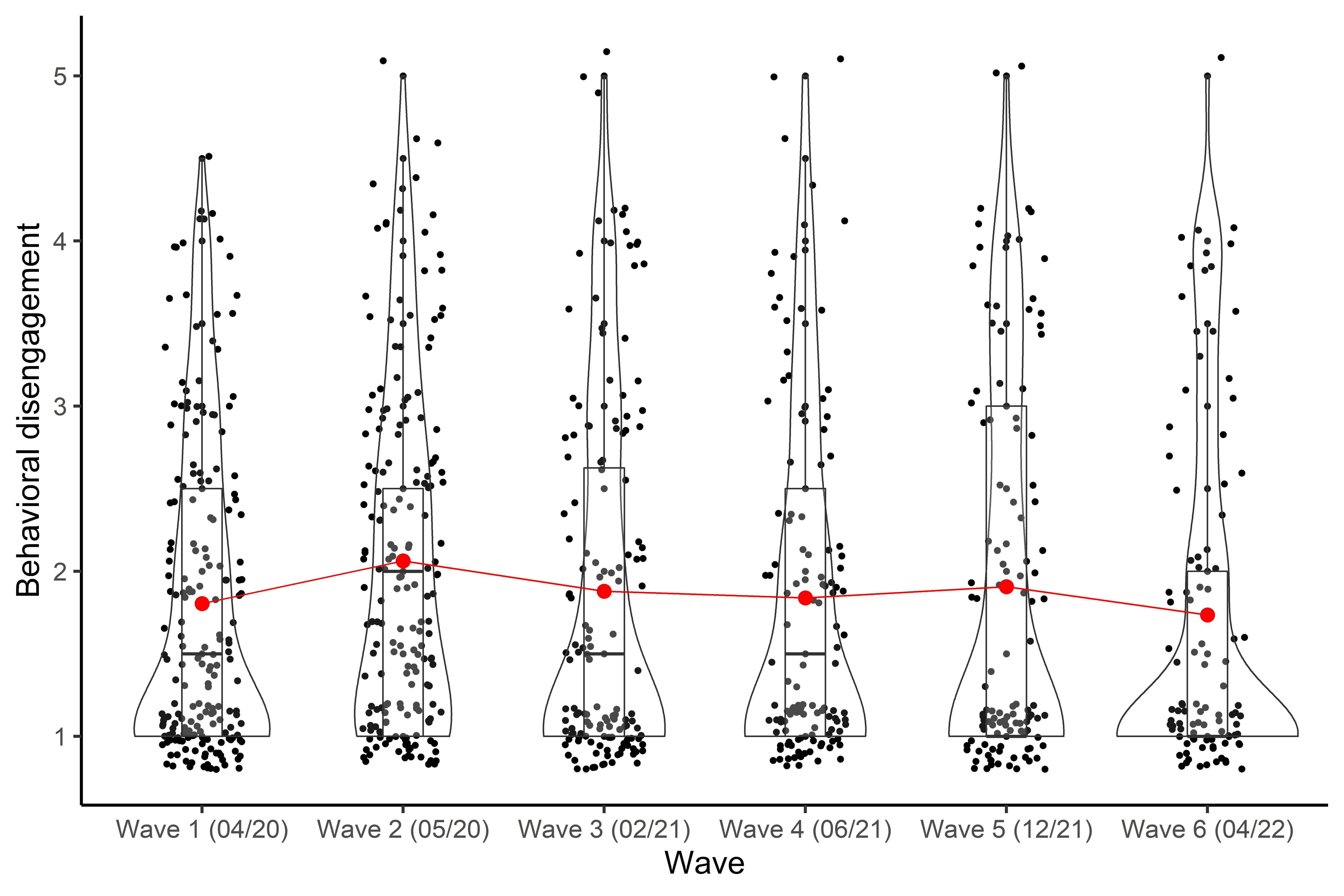} 
\caption{Behavioral disengagement across time. The red line displays the trend over time, whereas the box at each time point shows the range in which the middle 50\% of the data falls. Responses were given on a 5-point scale, with 1 being the lowest possible score and 5 being the highest possible score.}
\label{fig:bd}
\vspace{1cm}
\end{figure} 

\begin{figure}[h]
\centering
\includegraphics[width=0.8\linewidth]{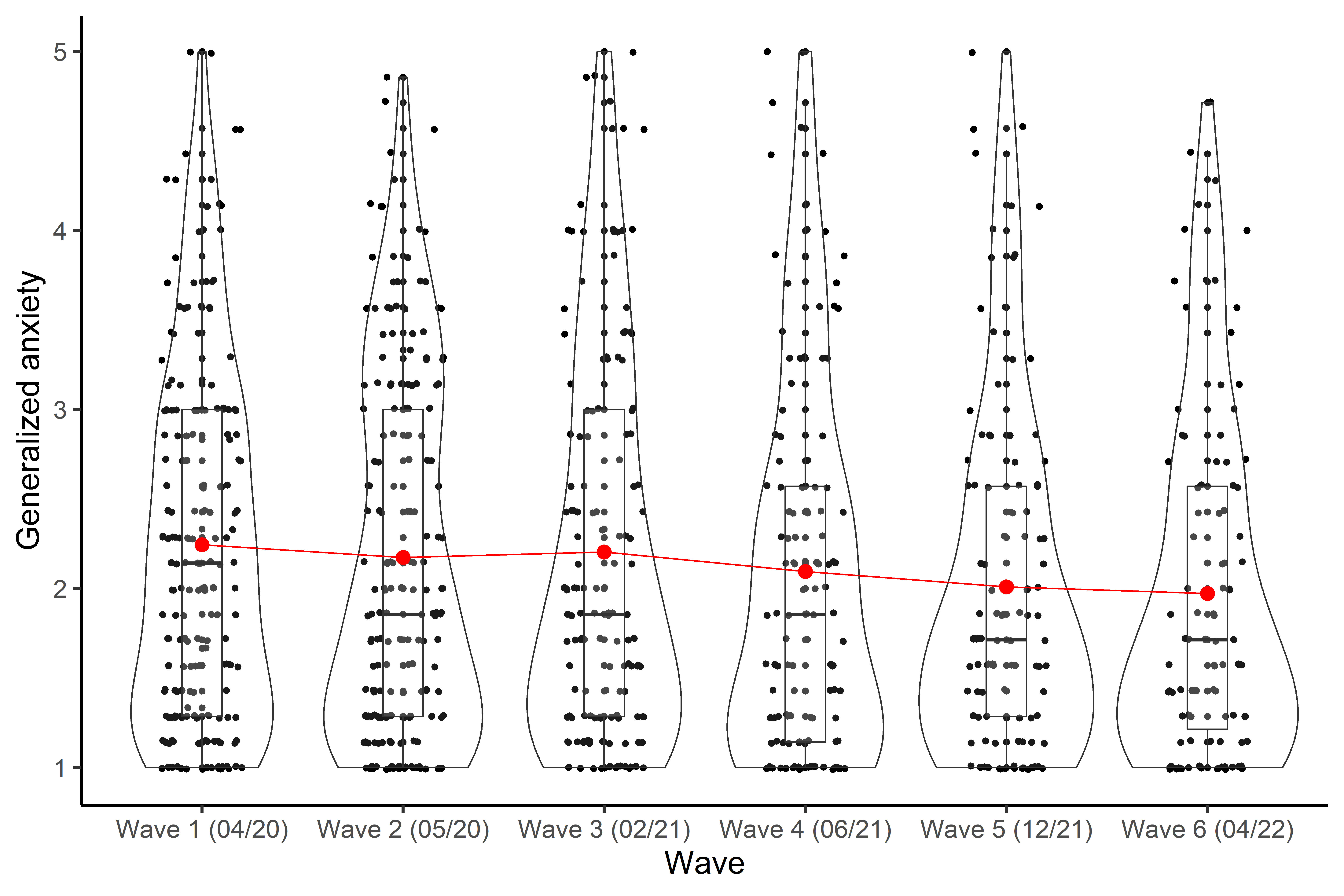} 
\caption{Generalized anxiety across time. The red line displays the trend over time, whereas the box at each time point shows the range in which the middle 50\% of the data falls. Responses were given on a 5-point scale, with 1 being the lowest possible score and 5 being the highest possible score.}
\label{fig:ga}
\end{figure}

\begin{figure}[h]
\centering
\includegraphics[width=0.8\linewidth]{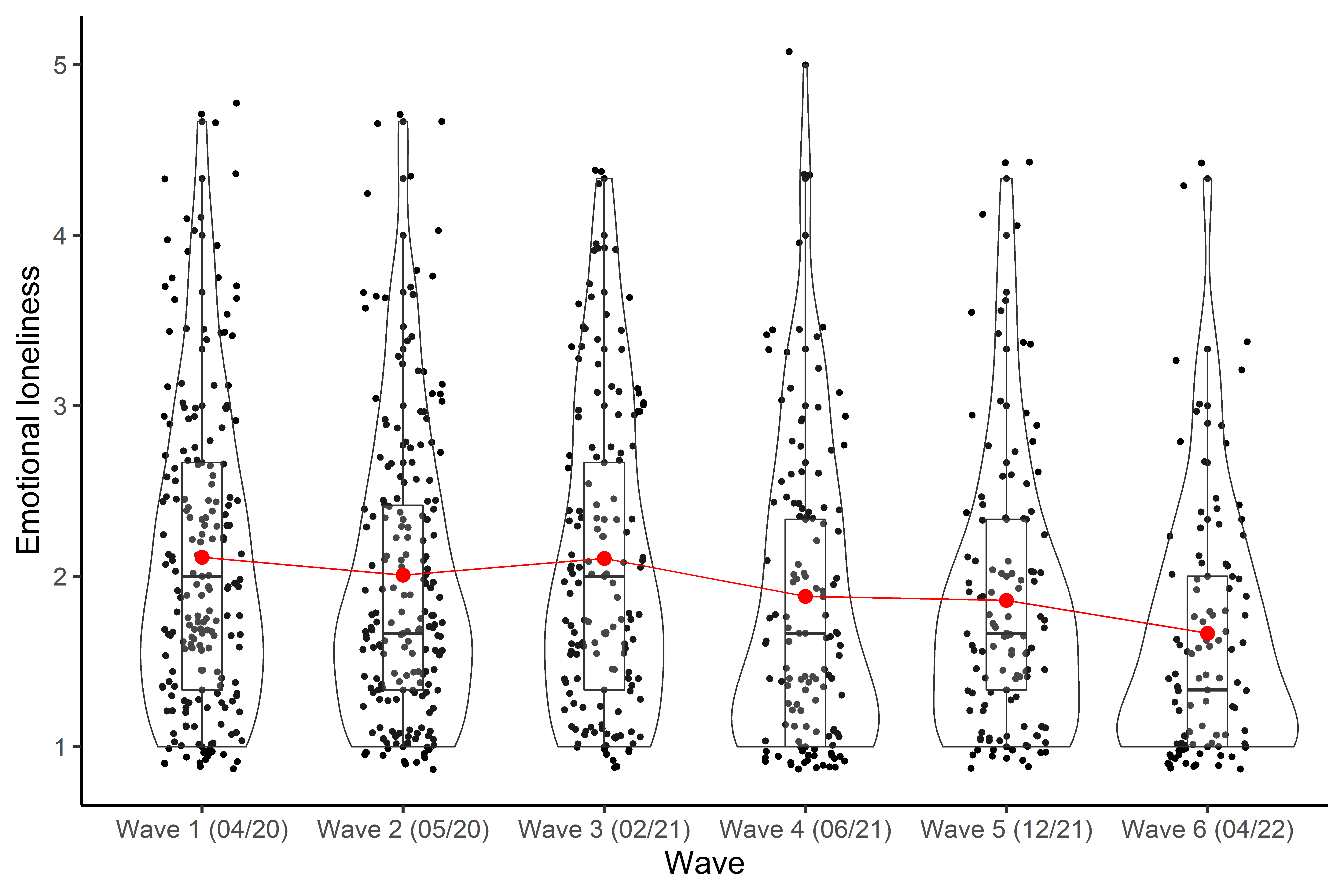} 
\caption{Emotional loneliness across time. The red line displays the trend over time, whereas the box at each time point shows the range in which the middle 50\% of the data falls. Responses were given on a 5-point scale, with 1 being the lowest possible score and 5 being the highest possible score.}
\label{fig:el}
\end{figure}

\begin{figure}[h]
\centering
\includegraphics[width=0.8\linewidth]{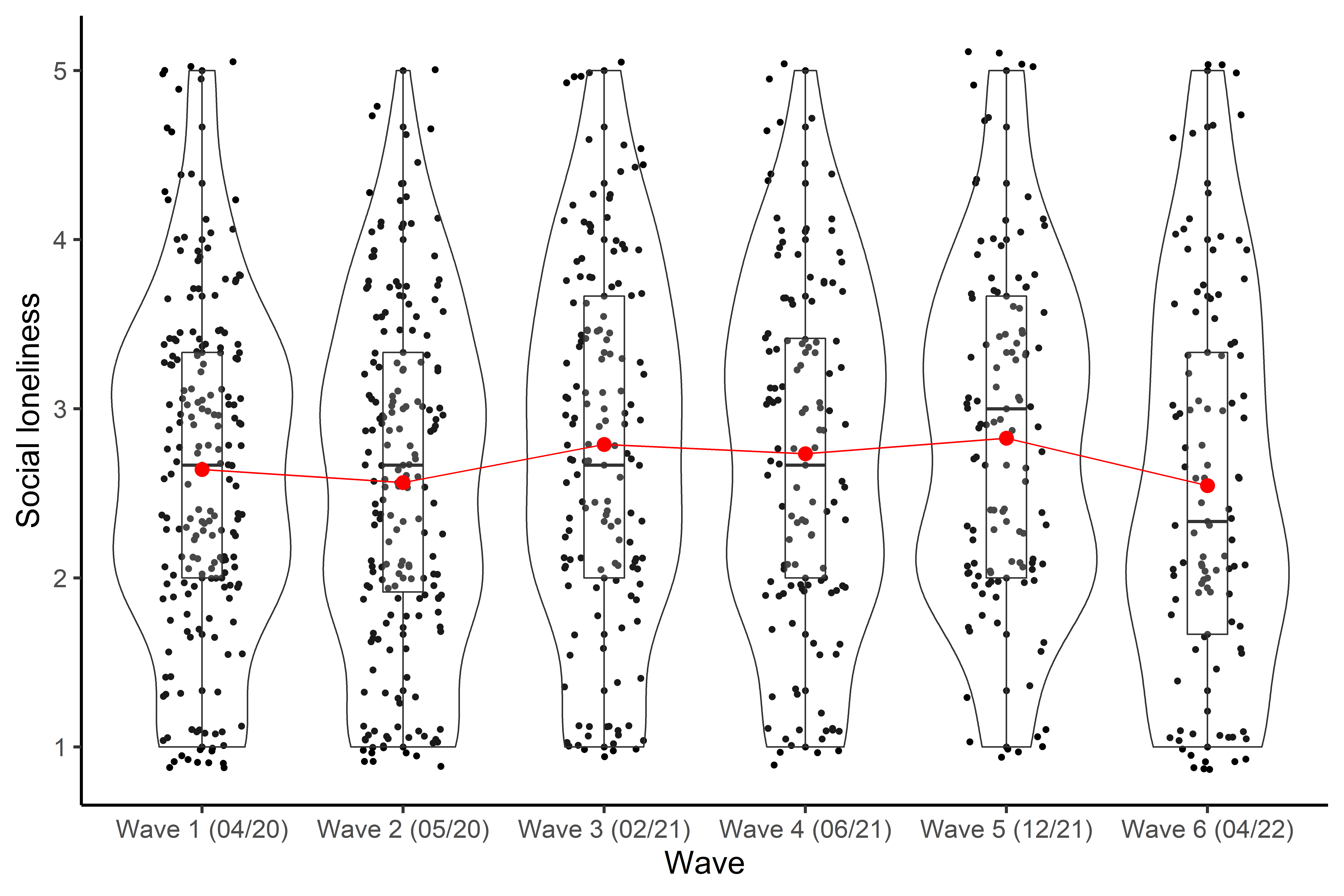} 
\caption{Social loneliness across time. The red line displays the trend over time, whereas the box at each time point shows the range in which the middle 50\% of the data falls. Responses were given on a 5-point scale, with 1 being the lowest possible score and 5 being the highest possible score.}
\label{fig:sl}
\end{figure}

\begin{figure}[h]
\centering
\includegraphics[width=0.8\linewidth]{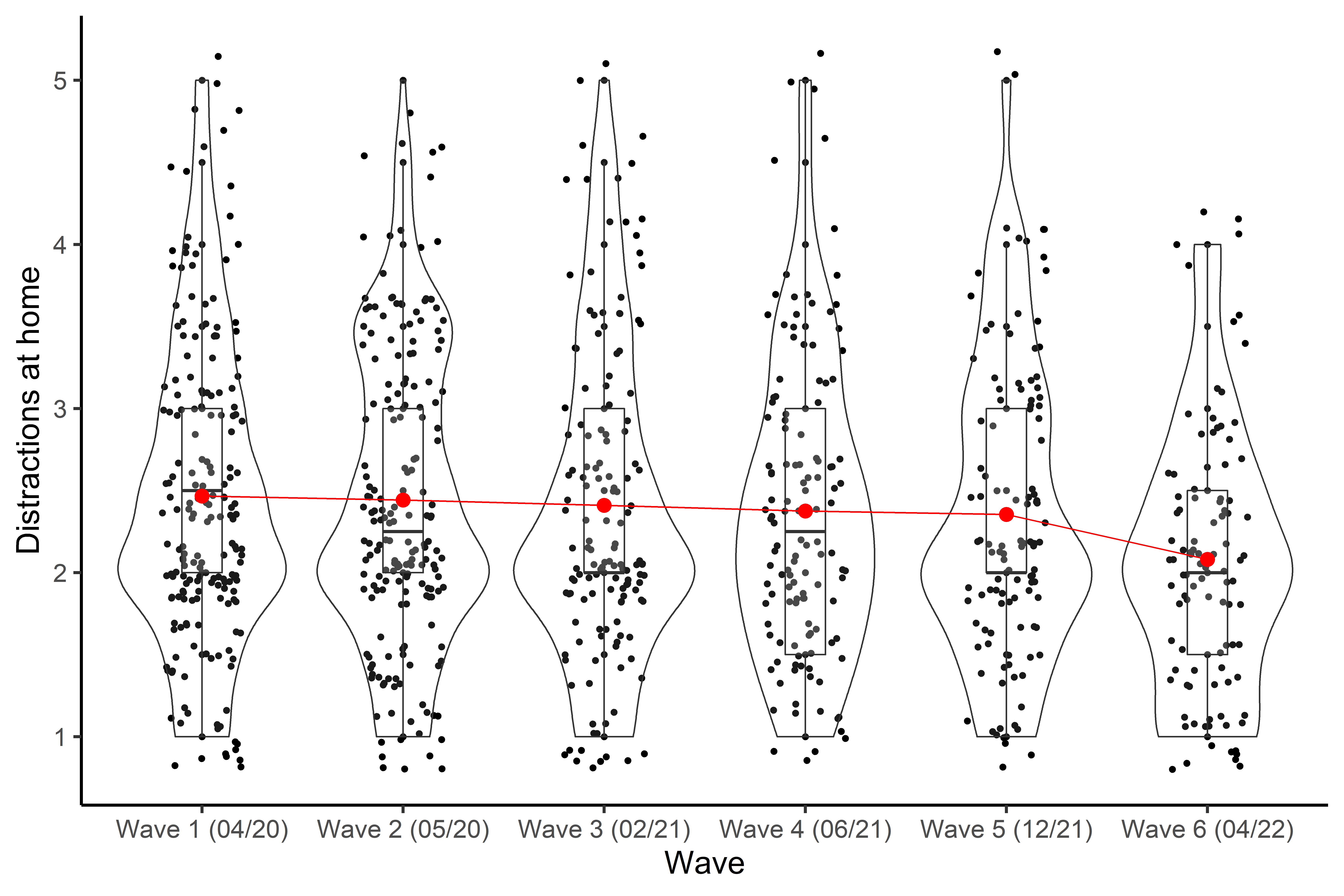} 
\caption{Distractions at home across time. The red line displays the trend over time, whereas the box at each time point shows the range in which the middle 50\% of the data falls. Responses were given on a 5-point scale, with 1 being the lowest possible score and 5 being the highest possible score.}
\label{fig:dh}
\end{figure}

\begin{figure}[h]
\centering
\includegraphics[width=0.8\linewidth]{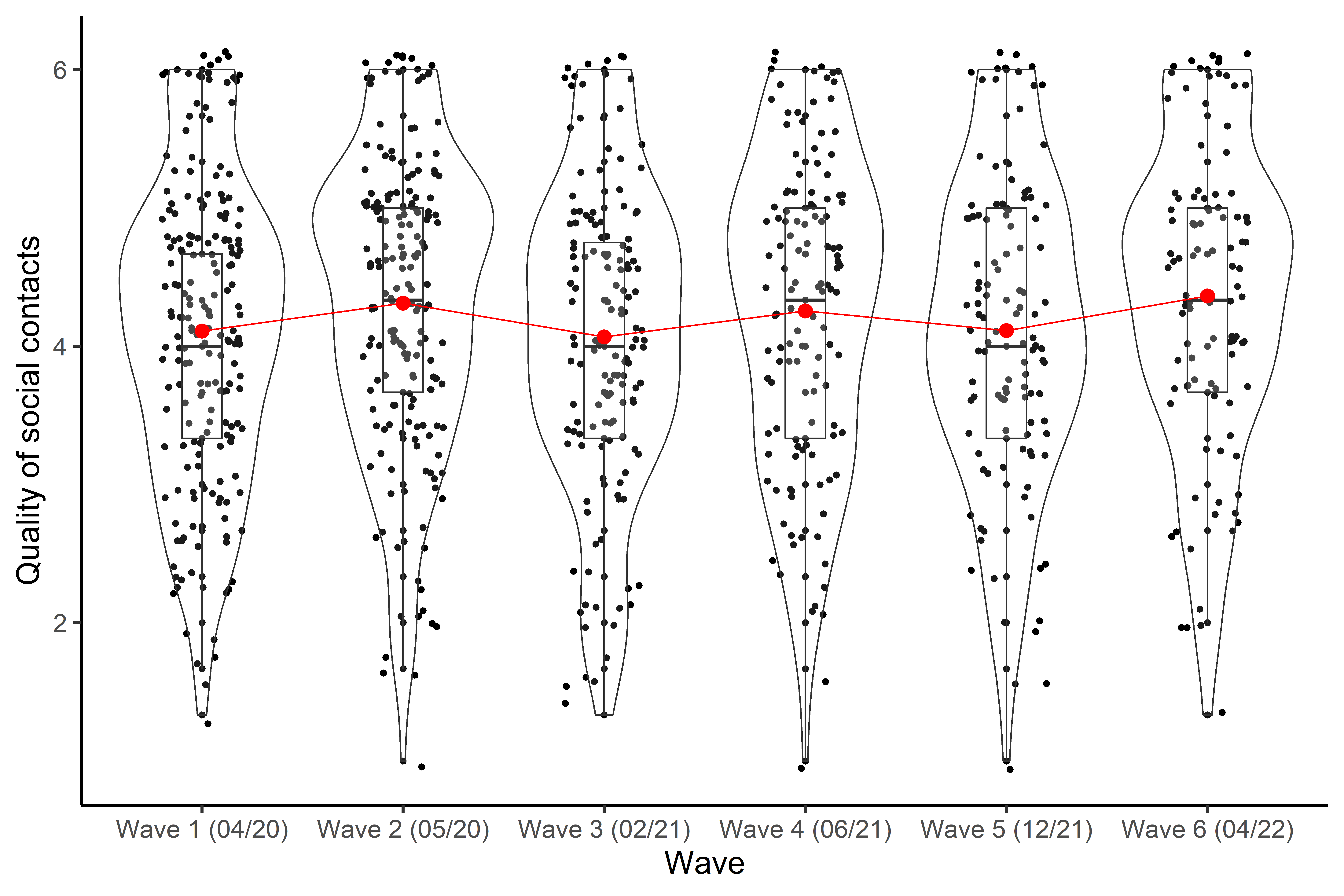}
\caption{Quality of social contacts across time. The red line displays the trend over time, whereas the box at each time point shows the range in which the middle 50\% of the data falls. Responses were given on a 6-point scale, with 1 being the lowest possible score and 5 being the highest possible score.}
\label{fig:sc}
\end{figure}

\end{appendix}

\end{document}